\documentclass[12pt,preprint]{aastex}  %--original for ApJ

\newcommand{\Alfven}{Alfv$\acute{e}$n}
\newcommand{\Alfvenic}{Alfv$\acute{e}$nic}

\shorttitle{Hot Region by Non-Equatorial MHD Shocks}
\shortauthors{Fukumura, Takahashi and Tsuruta}

\begin{document}

%\title{Properties of Fast Magnetohydrodynamic Shocks in Black Hole Magnetosphere [temporary title 1]}
%\title{Generation of Shocked Hot Plasma Region in a Black Hole Magnetosphere }
%\title{[1]Hot Region by Magnetohydrodynamic Shocks in Non-Equatorial Plasma Flows around a Black Hole}

\title{Magnetohydrodynamic Shocks in Non-Equatorial Plasma Flows around a Black Hole }

\author{Keigo Fukumura\altaffilmark{1}}
\affil{Department of Physics, Montana State University, Bozeman,
MT~59717;} \email{fukumura@physics.montana.edu}

\author{Masaaki Takahashi}
\affil{Department of Physics and Astronomy, Aichi University of
Education, Kariya, Aichi 448-8542, Japan;}
\email{takahasi@phyas.aichi-edu.ac.jp}

\and

\author{Sachiko Tsuruta}
\affil{Department of Physics, Montana State University, Bozeman,
MT~59717;} \email{uphst@gemini.msu.montana.edu}

\altaffiltext{1}{Present address: NASA/Goddard Space Flight Center,
Code 663, Greenbelt, MD 20771; fukumura@milkyway.gsfc.nasa.gov}

%\altaffiltext{2}{present email address:
%fukumura@milkyway.gsfc.nasa.gov}

%\author{\textsc{Keigo Fukumura}\altaffilmark{1,2}, \textsc{Masaaki
%Takahashi}\altaffilmark{3} \\
%\textsc{and} \\ \textsc{Sachiko Tsuruta}\altaffilmark{1} }
%
%\altaffiltext{1}{Department of Physics, Montana State University,
%Bozeman, MT~59717}
%
%\altaffiltext{2}{Current Address: NASA/Goddard Space Flight Center,
%Code 663, Greenbelt, MD 20771}
%
%\altaffiltext{3}{Department of Physics and Astronomy, Aichi
%University of Education, Kariya, Aichi 448-8542, Japan}

%\date{\today}

\begin{abstract}

We study magnetohydrodynamic (MHD) standing shocks in inflowing
plasmas in a black hole magnetosphere. Fast and intermediate shock
formation is explored in Schwarzschild and Kerr geometry to
illustrate general relativistic effects. We find that non-equatorial
standing MHD shocks are physically possible, creating a very hot
plasma region close to the event horizon. Shocked downstream plasmas
can be heated or magnetized depending on the values of various
magnetic field-aligned parameters. Then we may expect high-energy
thermal/nonthermal emissions from the shocked region. We present the
properties of non-equatorial MHD shocks and discuss the shocked
plasma region in the black hole magnetosphere. We also investigate
the effects of the poloidal magnetic field and the black hole spin
on the properties of shocks, and show that both effects can modify
the distribution of the shock front and shock strength. We find for
strong MHD shock formation that fast rotating magnetic fields are
necessary. The physics of non-equatorial MHD shocks in the black
hole magnetosphere could be very important when we are to construct
the central engine model of various astrophysical phenomena.

\end{abstract}


\section{Introduction}
We consider a theoretical implication of a possible link between the
conditions of magnetohydrodynamic (MHD) shocks and the resulting
shocked hot and/or strongly magnetized plasma region very close to
the black hole event horizon. In a series of our previous
investigations in the context of general relativity (see, Takahashi
et al. 2002, hereafter TRFT02; Rilett 2003, hereafter R03; Fukumura
2005; Takahashi et al. 2006, hereafter Paper I), it has been shown
that shock formation in MHD plasmas inflowing onto a black hole can
be a physically plausible mechanism for creating very hot (i.e.,
$T_i \gtrsim 10^{12}$ K for ions and $T_e \gtrsim 10^9$ K for
electrons) or strongly magnetized plasma regions, which possibly
could be associated with subsequent thermal/nonthermal high energy
emission. These previous studies are the extension of the earlier
works on hydrodynamical (HD) shock formation in black hole accretion
flows (e.g., Chakrabarti 1990; Lu et al. 1997; Lu \& Yuan 1998;
Fukumura \& Tsuruta 2004, hereafter FT04). These authors point out
that black hole rotation is also important for determining the shock
strength. The slow magnetosonic shock formation in relativistic
ingoing flows near a black hole was first explored by TRFT02. By
extending the work on shock formation in relativistic winds by
\citet{AC88}, these authors solved general relativistic jump
conditions in accreting plasmas. It was suggested that relativistic
slow/fast MHD shocks can significantly heat up the postshock plasma
(generation of hot regions), which may produce a sufficient amount
of high energy radiation. R03 systematically examined the types of
the preshock plasma mainly for slow MHD shocks. Fast MHD shock
formation in inflowing plasmas was first studied in Paper I where a
non-rotating black hole was primarily considered.

These studies were conducted mainly for the equatorial accretion
flows. Therefore, {\it a natural next step is to extend these
studies to two dimensional, non-equatorial flows.} Also,
non-equatorial shock-heated region would be attractive as a possible
high energy radiation source above an accretion disk in the central
engine of active galactic nuclei (AGNs). Therefore, we investigate
in this paper the polar angle dependence of various shock properties
(e.g., shock radius, compression ratio, number density,
magnetization, entropy generation and so on).

The goal of our current study is to suggest where in the parameter
space a MHD shock can possibly occur for specified black hole
magnetosphere models. That is, for a specified plasma source, we
hope that our model will be able to predict whether or not the MHD
standing shock formation is possible. Also, if the shock does
develop, then we hope that our model will be capable of suggesting
the physical nature of the shocked plasma. Solid understanding of
non-equatorial shock formation can be useful for comparing our
theoretical implications with future observations. Furthermore,
from observations of some Seyfert nuclei, it has been suggested
that the central black holes are rapidly rotating
\citep[e.g.,][]{Iwasawa96a,Iwasawa96b,Fab02,Wilms01}. Therefore,
we study the effects of black hole rotation as well, although in
this paper we will confine our attention to the black hole
rotation slower than that of the magnetic field lines of the
magnetosphere.

\begin{figure}[t]% ------------------------------------- Figure~1
    \centering
    \epsscale{0.5}
    \includegraphics[angle=0, width=3.5in]{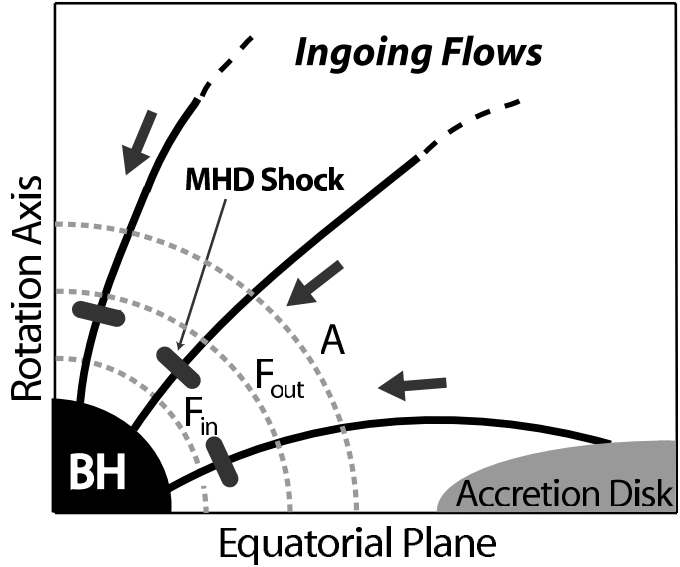}
    \caption{A schematic picture of standing MHD shock formation in black hole
    accretion flows where the poloidal magnetic field lines near the event horizon can be approximated
    to be radial (not drawn to scale). Note that non-equatorial ingoing MHD plasmas,
    originating from some plasma sources, are made possible
    in the presence of the poloidal magnetic fields. The \Alfven~point, outer and inner fast magnetosonic points
    (dotted curves) are labelled by A, F$_{\rm out}$ and F$_{\rm in}$, respectively. } \label{fig:fig1}
\end{figure}

In a black hole magnetosphere, the magnetic field geometry should in
principle be described by the trans-field force balance equation
(the Grad-Shafranov equation), which describes the mutual
interaction between the frozen-in plasma and the surrounding global
magnetic fields. As the plasma generates the toroidal/poloidal
current distribution, the poloidal/toroidal magnetic fields are
produced and the magnetosphere is constructed. Although solving this
trans-field balance equation is an extremely difficult problem,
there have been some attempts in the past to study the steady-state
magnetospheric configuration around black holes (see, for instance,
Mobarry \& Loveless 1986 for Schwarzschild geometry and Camenzind
1987 for Minkowski geometry). \citet{NTT91} extended these studies
to Kerr geometry in order to analytically study a rotating black
hole magnetosphere, and found dipole-like fields \citep[see
Figure~3b in][]{NTT91}. \citet{TT01} solved a vacuum magnetosphere
with a thin equatorial disk and obtained a dipole-like geometry in
the disk region and a uniform field geometry along the rotational
axis. A similar magnetospheric structure was discussed by
\cite{Li02}. However, for the accreting plasma very close to the
horizon, it was found that {\it the magnetic field lines are almost
radial in the poloidal plane} (see, e.g., Hirotani et al. 1992;
Komissarov 2005; Uzdensky 2005). Therefore, in our current paper we
mostly adopt a conical geometry because we are interested in the
accreting plasma very close to the event horizon. However, we will
also explore the effect of different magnetic field geometries.
Figure~\ref{fig:fig1} schematically illustrates a magnetic field
geometry in the poloidal plane adopted in this paper. We explore the
possibility of MHD shock formation along the poloidal magnetic field
lines, which may illuminate the underlying accretion disk, with
possible application to the central engine of AGN in mind. Although
the plasma flows can originate from the accreting material in the
equator, the exact plasma source can be diverse, and we do not
specify them in this paper.

%This is due to the plasma inertia effects.

From the standpoint of recent MHD simulations, the past several
years have seen the long-awaited advance in numerical general
relativistic MHD simulations to study the geometrical nature of
the black hole magnetosphere and the dynamical properties of the
plasma emersed in the global magnetic fields. These dynamical
simulations will help to offer the means to predict the boundary
conditions and parameters governing the black hole magnetosphere.
For example, \citet{Koide00} explored the roles of the global
magnetic fields coupled with the accreting plasma where the shock
formation is found in the equator, while \citet{Hirose04} analyzed
the structure of various magnetic fields around a rotating black
hole. The astrophysical processes that influence the spin
evolution of black holes are studied by \citet{Gammie04}.
%---[ cut by MT ] who estimated a final spin rate of $a \sim 0.8-0.9$.
Through long time-evolved MHD simulations, \citet{McKinney04}
found five main subregions of the black hole magnetosphere in a
quasi-steady state at the final simulation time.
\citet{DeVilliers05} recently showed that the unbound outflows can
emerge self-consistently in the axial funnel region where the
large-scale magnetic field spontaneously arises. The structure of
large-scale Poynting-dominated jets is discussed by
\citet{McKinney06}. On the other hand, in order to gain further
physical insights, here we emphasize that {\it parallel more
analytic, steady-state investigations should be equally important,
and that is a major goal of our current studies}.

From X-ray observations of accretion-powered AGNs, particularly
Seyfert galaxies, it has become evident that accretion disks play a
crucial role in the central high energy activities
\citep[e.g.,][]{Nandra94,Pounds94}. Rapid instrumental progress of
recent X-ray observatories (i.e., {\it Chandra} and {\it
XMM-Newton}) allows us to further investigate the detailed dynamics
of accreting plasmas in the immediate environment of supermassive
black holes hosted in these AGNs. It has been widely accepted that
the gravitational energy of the plasma is viscously dissipated into
thermal energy in the course of accretion, producing the signature
of thermal emission in the observed spectra in many Seyfert
galaxies. Furthermore, the presence of the power-law X-ray continuum
component suggests the existence of magnetic fields in/around the
accretion disks \citep[see, e.g.,][for an AGN model]{Haardt91}.
Therefore, in order to better understand the nature of the accreting
magnetized plasma near a black hole, it is important to consider the
role of both general relativity and magnetic fields. \citet{Meier04}
proposed that the inner part of the ingoing flows may enter a
magnetically-dominated, magnetosphere-like phase in some Galactic
black hole candidates (GBHCs) (e.g., GRS~1915+105) and perhaps some
low-luminosity AGNs (e.g., NGC~6251) as well. His model is based on
a strong, large-scale magnetic field structure around a black hole.
%-----------[ cut by Takahashi ]
% In the presence of such a large-scale magnetosphere,
% the rotational energy of a rapidly-rotating Kerr black hole can be
% extracted via the BZ-process \citep{BZ77} when the black hole is
% directly coupled to the surrounding disk through the magnetic
% fields \citep[e.g.,][]{Li02a}.
%------------//
These issues point to the importance of investigating the basic
physics of a {\it high energy source} in the black hole
magnetosphere. One motivation for our current paper is thus to
provide a scenario, in terms of MHD shock formation in inflowing
plasmas, that generates hot or magnetized plasma regions very
close (within a few gravitational radii) to a black hole.
%We stress here that the model in the series of our work on MHD
%shock formation is based on the presence of {\it global,
%well-ordered, poloidal magnetic fields} that can couple to both
%the black hole and plasma sources. Our ingoing plasma is assumed
%to originate from some plasma sources although we will not try to
%explain the plasma sources in this paper.

The structure of this paper is as follows. In \S2 we briefly
summarize our previous work on MHD shock formation in a rotating
black hole magnetosphere and describe the trans-magnetosonic
property of the plasma. We also show general relativistic
adiabatic MHD shock conditions for the possible parameter space.
The parameter dependence of the shock-related quantities is
explored in \S3, where we study the nature of our shock solutions
and shock-included global solutions. We particularly focus on the
non-equatorial MHD shock formation, by examining the polar angle
dependence and place constraints on possible allowed shock regions
in the parameter space. In \S4 we discuss our results in terms of
a black hole-plasma system coupled to a global magnetic field
lines. Brief summary and concluding remarks are given in the last
section, \S5.

\section{Assumptions \& Basic Equations}

\subsection{Black Hole Magnetosphere and MHD Accreting Flows }
We consider stationary and axisymmetric ideal MHD accretion flows
in Kerr geometry. The background metric is written by the
Boyer-Lindquist coordinates with the $c=G=1$ unit,
\begin{eqnarray}
   ds^2 &=& \left( 1-\frac{2mr}{\Sigma} \right) dt^2
        + \frac{4amr\sin^2\theta}{\Sigma} dt d\phi \nonumber \\
       & &  - \frac{A\sin^2\theta}{\Sigma} d\phi^2
            - \frac{\Sigma}{\Delta} dr^2 - \Sigma d\theta^2  \ ,
\end{eqnarray}
where $\Delta \equiv r^2 -2mr +a^2 $, $\Sigma \equiv r^2
+a^2\cos^2\theta$, $A\equiv (r^2+a^2)^2-a^2\Delta \sin^2\theta$,
and $m$ and $a$ denote the mass and angular momentum per unit mass
of the black hole, respectively. We take $m=1$ throughout this
paper. The black hole event horizon is $r_H \equiv
1+\sqrt{1-a^2}$.

In the context of general relativistic ideal MHD, the basic
equations governing plasmas consist of: (1) particle number
conservation law: $(n u^\alpha)_{;\alpha}$ where $n$ is the proper
particle number density and $u^\alpha$ is the plasma
four-velocity, (2) equation of motion:
$T^{\alpha\beta}_{~~;\beta}=0$ where $T^{\alpha \beta}$ is the
energy-momentum tensor for plasmas and (3) ideal MHD condition:
$u^\beta F_{\alpha \beta}=0$ where $F_{\alpha \beta}$ is the
electromagnetic field tensor. We also denote the poloidal plasma
velocity as $u_p^2 \equiv -(u^r u_r+u^\theta u_\theta)$. The
energy-momentum tensor $T^{\alpha \beta}$ is given by
\begin{eqnarray}
T^{\alpha \beta} \equiv n \mu u^\alpha u^\beta -P g^{\alpha \beta}
+ \frac{1}{4 \pi} \left(F^{\alpha \gamma} F_{\
\gamma}^{\beta}+\frac{1}{4} g^{\alpha \beta} F^2 \right) \ ,
\label{eq:plasma-tensor}
\end{eqnarray}
where $F^2 \equiv F_{\mu \nu} F^{\mu \nu}$ and $\mu=(\rho+P)/n$ is
the relativistic enthalpy, $P$ is the thermal gas pressure and
$\rho$ is the total energy density.

%In the Boyer-Lindquist coordinates, the toroidal component of the
%magnetic field seen by a distant observer is defined by $B_\phi
%\equiv (\Delta/\Sigma) F_{\theta r}$.

%Here, the expression for the toroidal component of magnetic field
%$B_\phi$ can be reduced to $B_\phi = -4 \pi \eta E \rho_w f$ where
%$\rho^2_w \equiv g^2_{t\phi}-g_{tt} g_{\phi \phi}$.

For a stationary and axially symmetric ideal MHD plasma, there
exist five conserved quantities along a field line: total energy
$E$ and angular momentum $L$ of the plasma, angular velocity of
the magnetic field line $\Omega_F$, particle flux per magnetic
flux $\eta$ and entropy $S$ \citep[see][R03]{Camenzind86}. The
total energy and angular momentum of the plasma are
\begin{eqnarray}
E &\equiv& \mu u_t -\frac{\Omega_F B_\phi}{4 \pi \eta}
\label{eq:E}
\ , \\
L &\equiv& -\mu u_\phi - \frac{B_\phi}{4 \pi \eta} \label{eq:L} \
,
\end{eqnarray}
where $\Omega_F \equiv -F_{tr}/F_{\phi r}=-F_{t \theta}/F_{\phi
\theta}$ \citep[see][]{Bekenstein78} and $\eta \equiv n u_p/B_p$.

In the Boyer-Lindquist coordinates, the toroidal component of the
magnetic field seen by a distant observer is defined by $B_\phi
\equiv (\Delta/\Sigma) F_{\theta r}$, and the poloidal one is
defined as $B^2_p \equiv - [g^{rr} (^\ast F_{rt})^2 + g^{\theta
\theta} (^\ast F_{\theta t})^2 ]G^{-2}_t$ where $^\ast F_{\alpha
\beta} \equiv (1/2) \sqrt{-g} \epsilon_{\alpha \beta \gamma
\delta} F^{\gamma \delta}$ is the tensor dual to $F_{\alpha
\beta}$ and $G_t \equiv g_{tt} + g_{t \phi} \Omega_F$. The
dynamical timescale of accretion is assumed to be much shorter
than diffusion timescale so that the adiabatic prescription is
used for the infalling plasma as $P = K \rho_0^\Gamma$, where
$\Gamma$ is the adiabatic index, $K$ is related to entropy $S$.
$\rho_0 = m_p n$ is the rest-mass density [i.e., $\rho = \rho_0 +
P/(\Gamma-1)$], and $m_p$ is the particle mass. We introduce the
so called entropy-related mass-accretion rate
\citep[e.g.,][]{Cha90} as
\begin{eqnarray}
\dot{\cal{M}} \equiv m_p \eta K^N \label{eq:Mdot} \ ,
\end{eqnarray}
the polytropic index $N$ is given by $\Gamma=1+1/N$. Note that
$\dot{\cal{M}} = 0$ for the cold flow limit (because of $K = 0$).

The poloidal equation is given by \citep[see, e.g.,][]{TNTT90}
\begin{eqnarray}
 \mu^2 (1+u_p^2) = E^2 \left[(\alpha-2M^2)f^2 -k \right]  \ ,
                                              \label{eq:poloidal}
\end{eqnarray}
with $f \equiv -(G_\phi + G_t \tilde{L})/[\rho_w (M^2-\alpha)] $,
$\alpha \equiv g_{tt}+2g_{t\phi}\Omega_F+g_{\phi\phi}\Omega_F^2$,
$ k \equiv (g_{\phi\phi} + 2g_{t\phi} \tilde{L} + g_{tt}
\tilde{L}^2)/\rho_w^2$, $\tilde{L} \equiv L/E$, $\rho^2_w \equiv
g^2_{t\phi}-g_{tt} g_{\phi \phi}$ and $G_\phi \equiv
g_{t\phi}+g_{\phi\phi}\Omega_F=g_{\phi\phi}(\Omega_F-\omega)$,
where $\omega\equiv -g_{t\phi}/g_{\phi\phi}$ is the angular
velocity of a zero angular momentum observer (ZAMO). The
relativistic Alfv\'{e}n Mach number $M$ is defined as
\begin{equation}
 M^2\equiv \frac{4\pi\mu n u_p^2}{B_p^2}     %= \frac{u_p}{\Bp} \ ,
         = \frac{4\pi\mu\eta u_p}{B_p}  \ .  \label{eq:Mach}
\end{equation}
The toroidal magnetic field $B_\phi$ can be expressed as
$B_\phi = -4 \pi \eta E \rho_w f$.

Any physical MHD accreting plasma ejected from the plasma source
with a small poloidal velocity onto a black hole must become fast
magnetosonic at the event horizon, going through three
magnetosonic critical points: a slow magnetosonic point
(S:~$r=r_{\rm S}$ where $u_p=u_{\rm SW}$), the \Alfven ~point
(A:~$r=r_{\rm A}$ where $u_p=u_{\rm AW}$) and a fast magnetosonic
point (F:~$r=r_{\rm F}$ where $u_p=u_{\rm FW}$). At each
magnetosonic point the plasma speed is equal to each magnetosonic
wave speed. Here, $u_{\rm SW},u_{\rm AW}$ and $u_{\rm FW}$ are the
slow, the \Alfven ~and the fast magnetosonic wave speed,
respectively \citep[see][for their definitions]{TNTT90}. When the
preshock flow is cold (i.e., $P=0$ or $\mu=m_p$), the ejected
plasma is super-slow magnetosonic; the slow magnetosonic point
will be absent in the cold flow solution.

The function $f$ becomes zero when the numerator of the expression
of $f$, $G_\phi + G_t \tilde{L}$, is zero. There are one or two
radii saisfying this condition and we will denote this radius
($r=r_{\rm A}$) the ``\Alfven ~radius''. When the plasma poloidal
speed becomes equal to the \Alfven ~speed ($u_p = u_{\rm AW}$ or
$M^2 = \alpha$) at this radius, this point is called the ``\Alfven
~point''. Otherwise, it is termed the ``anchor point'' where the
toroidal component of the magnetic field $B_\phi$ is zero ($M^2
\ne \alpha$). In the next section, let us apply MHD shock
conditions to these accreting plasma flows.

%The \Alfven ~point is dynamically important while the anchor point
%only has a geometrical meaning in the toroidal magnetic field
%configuration.

%Figure~\ref{fig:trans-magnetosonic} shows an example of the
%trans-magnetosonic accreting plasma passing through the
%magnetosonic points before reaching the event horizon (H:$r=r_H$).

\subsection{General Relativistic MHD Shock Conditions}
In TRFT02 and Paper I, we discussed in detail general relativistic
Rankine-Hugoniot shock conditions in Kerr geometry that apply to
accreting plasmas: particle number conservation $[n u^\alpha]_{\rm
sh} ~\ell_\alpha=0$, energy and angular momentum conservation
$[T^{\alpha \beta}]_{\rm sh} ~\ell_\alpha=0$ and magnetic flux
conversation $[^{\ast} F^{\alpha \beta}]_{\rm sh} \ell_\alpha=0$,
where $\ell_\alpha$ is the unit vector normal to the shock front.
The square brackets denote the difference between the values of a
quantity on the two sides of the shock front.

The poloidal magnetic field in our assumption is explicitly given
by $B_p(r) = C/(\sqrt{\Delta \Sigma} ~r^{\delta})$ where $C$ is a
constant.
%---[ cut by MT ]
% the constant $C$ corresponds to the field strength at a plasma
% source (i.e., an injection point).
%---//
To investigate the effects of the non-radial poloidal magnetic
field, the poloidal field is parameterized by $\delta$: $\delta=0$
for a purely conical field and $\delta>0$ for an outward diverging
field. The conical magnetic field with $\delta=0$ is adopted in
most of our computations unless otherwise stated. In the
following, for simplicity we set the shock front to be normal to
the magnetic field lines, $\ell_\alpha=(0,1,0,0)$. These
conservation laws lead to the following jump conditions:
\begin{eqnarray}
n_1 u^r_1 = n_2 u^r_2 \label{eq:number} \ ,
\end{eqnarray}
\begin{eqnarray}
\left(\frac{1}{\sqrt{\Delta \Sigma} ~r^{\delta}} \right)^2 M^2_1 +
\frac{\hat{\mu}_1 \hat{\eta}}{M^2_1} \left[\frac{\hat{\eta}
\left(\hat{\mu}_1-1 \right)}{1+N} \right] +
\frac{1}{2} (\hat{\eta} \hat{E} )^2 f^2_1 & & \nonumber \\
= \left(\frac{1}{\sqrt{\Delta \Sigma} ~r^{\delta}} \right)^2 M^2_2
+  \frac{\hat{\mu}_2 \hat{\eta}}{M^2_2} \left[\frac{\hat{\eta}
\left(\hat{\mu}_2-1 \right)}{1+N} \right] + \frac{1}{2}
(\hat{\eta} \hat{E})^2 f^2_2 \ , \label{eq:shock-condition}
\end{eqnarray}
and the relation:
\begin{eqnarray}
\left(\hat{\mu} \hat{\eta} \right)^2 &=& (\hat{\eta} \hat{E})^2
\left[\left(\alpha-2M^2 \right)f^2 -k \right] - \left(
\frac{1}{\sqrt{\Delta \Sigma} ~r^{\delta}}\right)^2 M^4_2 \ .
\label{eq:mu12}
\end{eqnarray}
where $\hat{\mu} \equiv \mu / m_p$, $\hat{E} \equiv E/m_p$ and
$\hat{\eta} \equiv 4 \pi \eta m_p/C$. The subscripts ``1'' and
``2'' respectively denote the preshock and postshock quantities.
The field-aligned conserved quantities
($E,\tilde{L},\Omega_F,\eta$) are continuous across a MHD shock.
Then, our entropy-related accretion rate can be expressed as:
\begin{eqnarray}
\dot{\cal{M}} &=& \frac{C}{m_p} \frac{M^2}{\hat{\mu} \hat{\eta}}
\left( \frac{\hat{\mu} -1}{1+N} \right)^N \ . \label{eq:Mdot-sk}
\end{eqnarray}
Note that $\dot{\cal{M}}$ is continuous in shock-free flow
regions, but it is discontinuous across a shock ($\dot{\cal{M}}_1
< \dot{\cal{M}}_2$) because of the entropy generation,
%---[ cut by MT ]
% $K$,
%---//
where the adiabatic shock is related to extremely inefficient
cooling processes. That is, most of the heat generated at the
shock front is carried away (advected) with the accreting plasma.

It should be noted from the shock condition expressed by
equations~(\ref{eq:shock-condition}) and (\ref{eq:mu12}) that the
plasma energy $\hat{E}$ is coupled with the parameter $\hat{\eta}$
as $\hat{\eta} \hat{E}$, while the angular momentum $L$ is always
coupled with $E$ as $\tilde{L} \equiv L/E$ through functions $f$
and $k$. Also, the enthalpy $\hat{\mu}$ in
equation~(\ref{eq:shock-condition}) frequently appears with
$\hat{\eta}$ as $\hat{\mu} \hat{\eta}$, which can be expressed in
terms of parameters $\hat{\eta} \hat{E}$, $\tilde{L}$ and
$\Omega_F$ through equation~(\ref{eq:mu12}). When the downstream
plasma becomes extremely hot (i.e., large $\hat{\mu}_2 \gg 1$),
the coupled parameter, $\hat{\eta} \hat{E}$, acts as a controlling
parameter for determining the shock solutions. The
equation~(\ref{eq:shock-condition}) also shows that the loss of
kinetic energy leads to the gain of the (toroidal) magnetic and
thermal energy across the fast MHD shocks.
%---[ cut by MT ]
% In the shock condition
% given by equation~(\ref{eq:shock-condition}) we expect weak shocks
% as $\hat{\eta} \hat{E} \rightarrow 0$ because then $\hat{\mu}_2
% \rightarrow \hat{\mu}_1$ (small heat generation) and $M_2
% \rightarrow M_1$ (small jump in the Mach number) for both sides to
% be identical to one another.
%---//
Thus, in our computations, we will
use the coupled parameter $\hat{\eta} \hat{E}$.

We now introduce some shock-related quantities that are useful for
examining the properties of MHD shocks. First, let us define the
magnetization parameter $\sigma$, which denotes the ratio of the
Poynting flux to the net mass-energy flux of the accreting plasma
seen by the ZAMO (see TRFT02). From the ZAMO's standpoint,
$\sigma$ can be written as
\begin{eqnarray}
\sigma \equiv \frac{B_{\phi} g_{\phi \phi} (\Omega_F-\omega)}{4
\pi \eta \mu u^t \rho^2_w} = - \frac{\tilde{e}-\alpha
\tilde{h}}{\tilde{e}-M^2 \tilde{h}} \ \label{eq:sigma},
\end{eqnarray}
where $\tilde{h} \equiv g^{tt} (1-\omega \tilde{L})$ and
$\tilde{e} \equiv 1-\tilde{L}\Omega_F$.

%--ST comment: define ZAMO if not done earlier!

%We also introduce a toroidal field amplification factor $q$ as
%\begin{eqnarray}
%q \equiv \frac{B_{\phi2}}{B_{\phi1}} =
%\frac{M^2_1-\alpha}{M^2_2-\alpha} \ \label{eq:q} .
%\end{eqnarray}
%Here, $q>1$ for the fast MHD shock, $0 \le q<1$ for the slow MHD
%shock, $q<0$ for the intermediate MHD shock.

The local strength of a shock is measured by the ratio of the
postshock to the preshock number density, $n_2/n_1$, as
\begin{eqnarray}
\lambda \equiv \frac{n_2}{n_1} \ .
\end{eqnarray}
Since any physical shocks that satisfy the second law of
thermodynamics must be compressible, we require $\lambda>1$ for
physically relevant shock formation. Then, the postshock plasma is
heated up. In order to quantitatively evaluate shock heating, we
furthermore define a dimensionless postshock plasma temperature
$\Theta$ as
\begin{eqnarray}
\Theta \equiv \frac{k_B T}{m_p} = \frac{\hat{\mu} -1}{1+N}
 \ , \label{eq:Theta}
\end{eqnarray}
assuming the equation of state for an ideal gas (i.e., $P=n k_B T$
where $k_B$ is the Boltzmann constant and $T$ is the plasma
temperature). Notice that $\Theta \ll 1$ for the cold flow
($P_1=0$ and $\mu_1=m_p$). Depending on the temperature $\Theta$,
the postshock flow, heated by the shock, may be capable of
emitting high energy thermal radiation, although in practice it is
also necessary to consider non-thermal radiation due to the
magnetic field.

To include the radiative effects on MHD flows, which means the
breakdown of ideal MHD approximation, we should formulate the
general relativistic version of radiative (non-ideal)
trans-magnetosonic flow solutions, where
$E,L,\eta,\Omega_F,\dot{\cal{M}}$ are not conserved along the
magnetic field line. However, the task is very complicated, and
that is beyond the scope of our present paper. Therefore, we will
not include the radiative effects on the postshock hot flows and
consider ideal MHD trans-magnetosonic flows.
%Lastly, for a computational purpose in the following sections, we
%define $\hat{E} \equiv E/m_p$ and $\hat{\eta} \equiv 4 \pi \eta
%m_p/C$.

\subsection{Shock-Included Trans-magnetosonic Accretion Flows}
In the following we consider the MHD shock formation around a
Schwarzschild and slowly-rotating Kerr black hole (i.e., $0 \le
\omega_H < \Omega_F$ where $\omega_H \equiv \omega(r_H)$, see also
Paper I). Although the black hole accretes the surrounding gas
inward by its strong gravity, there are several mechanisms in
general for decelerating the accreting gas: the centrifugal force
due to plasma's angular momentum, magnetic tension and magnetic
pressure. For accretion onto black holes a shock can be formed
through these obstacles. Detailed studies of this possibility will
be a focus of much of this section. We restrict ourselves to cold
preshock plasmas which are initially injected from a plasma source
(e.g., accretion disk/torus). We impose that the cold plasma starts
falling onto the black hole with small poloidal velocity (i.e., $u_p
< u_{\rm AW}$) from this injection point (I) which is located
somewhere between the outer light surface (L) and the \Alfven ~point
(A). The exact location of the injection point will be important
when considering the boundary conditions of the accreting plasma
that should be specified by some accretion disk model, but that is
beyond the scope of our present paper.

%----[ move from page 14   by MT ]
We should point out the rotational energy per total energy,
$\tilde{L} \Omega_F$, is one of the useful parameters for predicting
the degree of magnetization of the shocked plasma flow. In Paper I
we discussed that there is a finite range in the value for
$\tilde{L} \Omega_F$ in order for the multiple \Alfven~points to
exist for the $\Omega_F>\omega_H$ case; that is, $(\tilde{L}
\Omega_F)_{\rm min} < \tilde{L} \Omega_F < (\tilde{L} \Omega_F)_{\rm
max}$. As the value of $\tilde{L} \Omega_F$ becomes larger, the
plasma becomes more magnetized because in the energy equation
(\ref{eq:E}) the magnetic term (second term) becomes more effective
as $\tilde{L} \Omega_F$ increases.
%----

Here, we briefly describe a shock-included trans-magnetosonic
accretion flow. The poloidal equation~(\ref{eq:poloidal}) specifies
the velocity of a plasma flow for a given set of initial parameters.
When a MHD shock develops, both preshock and postshock flow
solutions must pass through appropriate critical points: that is,
the flow must pass through the fast magnetosonic points twice ({\it
outer} and {\it inner} fast magnetosonic points). As discussed
earlier, for a given set of conserved quantities, there exist two
\Alfven ~radii : {\it outer} and {\it inner} \Alfven ~radii for a
slowly-rotating black hole case. We consider that an upstream
preshock solution goes through the outer \Alfven ~radius while the
corresponding downstream postshock solution passes through the inner
\Alfven ~radius. The {\it outer} \Alfven ~radius is identified as
the \Alfven ~point (A) while the {\it inner} \Alfven ~radius
corresponds to the anchor point for the upstream solution. The
anchor point is located somewhere between the outer fast
magnetosonic point F$_{\rm out}$ and the fast MHD shock location.
When a shock forms inside the anchor point, the magnetic field line
is refracted away from the shock normal. Such a shock is called a
fast MHD shock. In fast MHD shocks, the preshock plasma must be
super-fast magnetosonic while the postshock plasma must be sub-fast
magnetosonic. This postshock plasma must pass through another fast
magnetosonic point (F) again before reaching the horizon (H) (i.e.,
I $\rightarrow$ A $\rightarrow$ F$_{\rm out}$ $\rightarrow$ Fast
Shock $\rightarrow$ F$_{\rm in}$ $\rightarrow$ H). When a shock
forms outside the anchor point, the magnetic field line is flipped
over across the shock normal. Such a shock is called an intermediate
MHD shock \citep[see, e.g.,][for its physical
significance]{Hada94,DeSterck00,DeSterck01}. In intermediate MHD
shocks, the preshock plasma must be super-\Alfvenic ~while the
postshock plasma must be sub-\Alfvenic. This postshock plasma must
pass through another \Alfven ~point (A) and a fast magnetosonic
point (F) subsequently again before reaching the horizon (H) (i.e.,
either I $\rightarrow$ A$_{\rm out}$ $\rightarrow$ Intermediate
Shock $\rightarrow$ A$_{\rm in}$ $\rightarrow$ F$_{\rm in}$
$\rightarrow$ H or I $\rightarrow$ A$_{\rm out}$ $\rightarrow$
F$_{\rm out}$ $\rightarrow$ Intermediate Shock $\rightarrow$ A$_{\rm
in}$ $\rightarrow$ F$_{\rm in}$ $\rightarrow$ H). When a shock
occurs right at the anchor point, it is called a ``switch-on shock''
for which the magnetic field line of the preshock flow is radial.
%---[ cut by MT ]
% because the toroidal field is switched on across the shock.
%---
The shock condition
(\ref{eq:shock-condition}) determines whether a MHD shock can form
for a given field-aligned parameter sets. In such a global
accretion solution with a MHD shock, the preshock cold plasma
[i.e., $\dot{\cal{M}}_1(r_{\rm sh},M^2_1)=0$] is connected to the
subsequent postshock hot plasma [i.e., $\dot{\cal{M}}_2(r_{\rm
sh},M^2_2)>0$] through the shock formation where the \Alfven~Mach
number jumps from $M_1$ to $M_2$ (see Paper I). In the next
section we systematically explore the parameter dependence of the
resulting MHD shocks by varying certain primary quantities.

\section{Numerical Results}
In this section, the shock-included trans-magnetosonic accretion
solutions are solved in wide ranges of field-aligned flow
parameters, where slowly-rotating black hole cases are considered;
that is, the magnetic field lines in the black hole magnetosphere
rotate faster than the black hole itself. In this case, there are
two \Alfven ~ radii on the $r$-$M^2$ plane. The locations are
specified by the two field-aligned parameters, namely $\Omega_F$
and $\tilde{L}$. These parameters $\Omega_F$ and $\tilde{L}$ have
the minimum and maximum values for allowing the \Alfven~ radii.
These values depend on the geometry of magnetic field lines. In
our assumed conical field configuration, these values depend on
the polar angle $\theta$. We must change the values of $\Omega_F$
and $\tilde{L}$ in the acceptable range in order to search for
physical solutions. AC88 and R03 both have suggested that
$\Gamma=4/3$ is not a good approximation even for large radial
velocity. In many fast shock cases, R03 has found that $\Gamma
\sim 3/2$ (i.e., $N \sim 2$). Following their claim, we choose
$\Gamma=3/2$ ($N=2$) in this paper. We have checked that this
choice is not qualitatively too critical for our end results.

As to the choice of field-aligned parameter sets for computations,
there are additional {\it three} restrictions for selecting our
parameters: regularity conditions at the {\it outer} and {\it inner}
fast magnetosonic points and the shock condition connecting preshock
and postshock trans-magnetosonic flows. Three out of five conserved
parameters are related by these restrictions. Consequently, for a
given black hole spin $a$ and the magnetic field line with polar
angle $\theta_{\rm sh}$, the shock location $r_{\rm sh}$ is
determined when the five parameters are selected in the acceptable
parameter range, and the corresponding shock quantities are
obtained. If we modify one of the parameters [i.e., the
field-aligned parameters ($E,L,\eta,\Omega_F$) + geometrical
parameters $(a,r_{\rm sh},\theta_{\rm sh},\delta)$], to find another
global shock solution, we must change at least one more
field-aligned parameter. Therefore, to clarify the $\theta_{\rm
sh}$-dependence of MHD shock solutions, we change one parameter at a
time in such a way that the degeneracy in shock solutions is
removed. This allows us to systematically examine each
parameter-dependence of the shock solutions. As explained in section
\S2.2, the shock condition in equation~(\ref{eq:shock-condition}) is
conveniently characterized by $\hat{\eta} \hat{E}$ and $\tilde{L}
\Omega_F$, and therefore in the rest of the computations we will
basically use $\hat{\eta} \hat{E}$ and $\tilde{L} \Omega_F$ as
control parameters.

\subsection{Non-equatorial MHD Shock Solutions}

\begin{figure}[h]% ------------------------------------- Figure~2
    \centering
    \epsscale{0.5}
    \includegraphics[angle=0, width=2.5in]{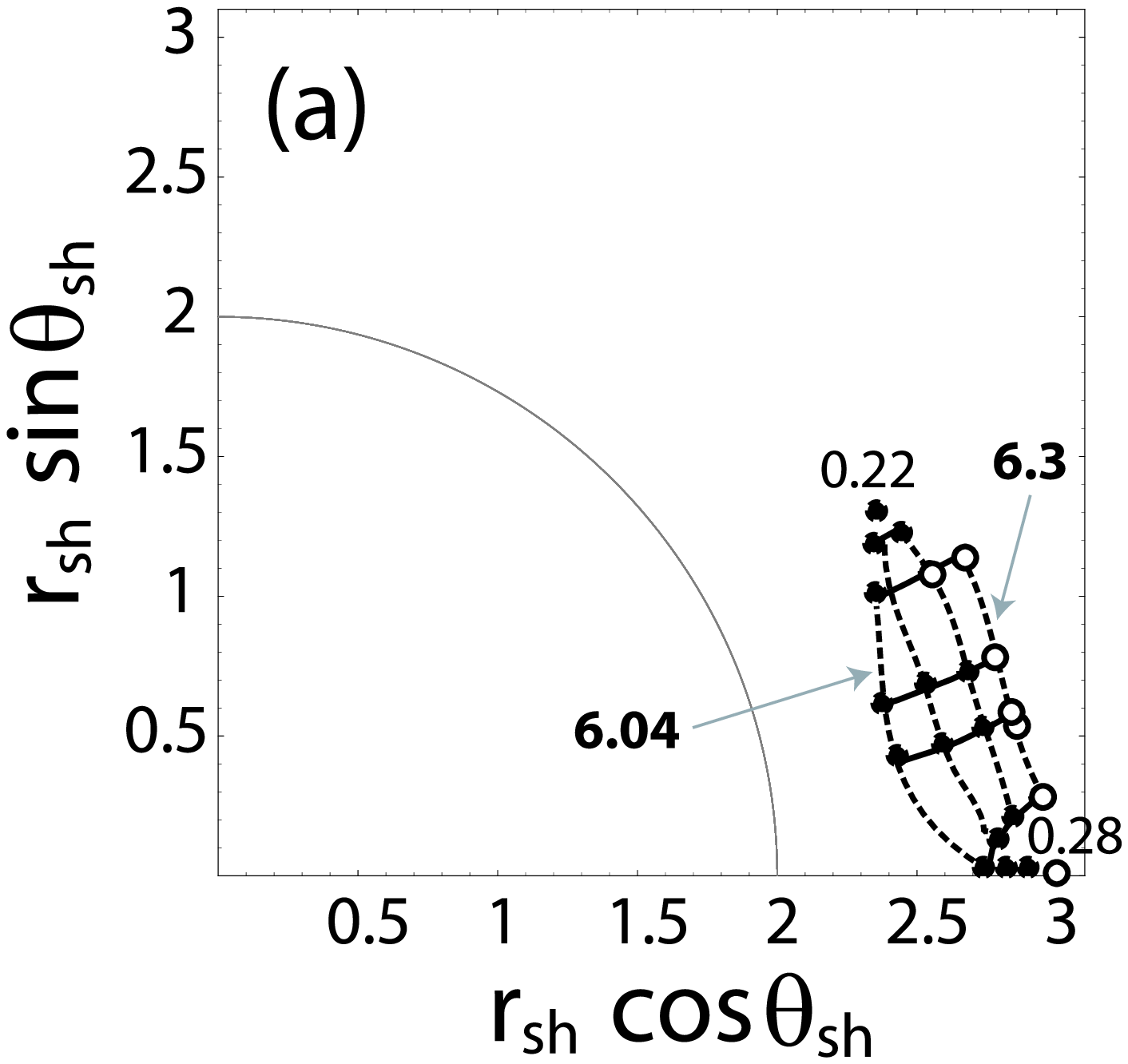}
    \includegraphics[angle=0, width=2.5in]{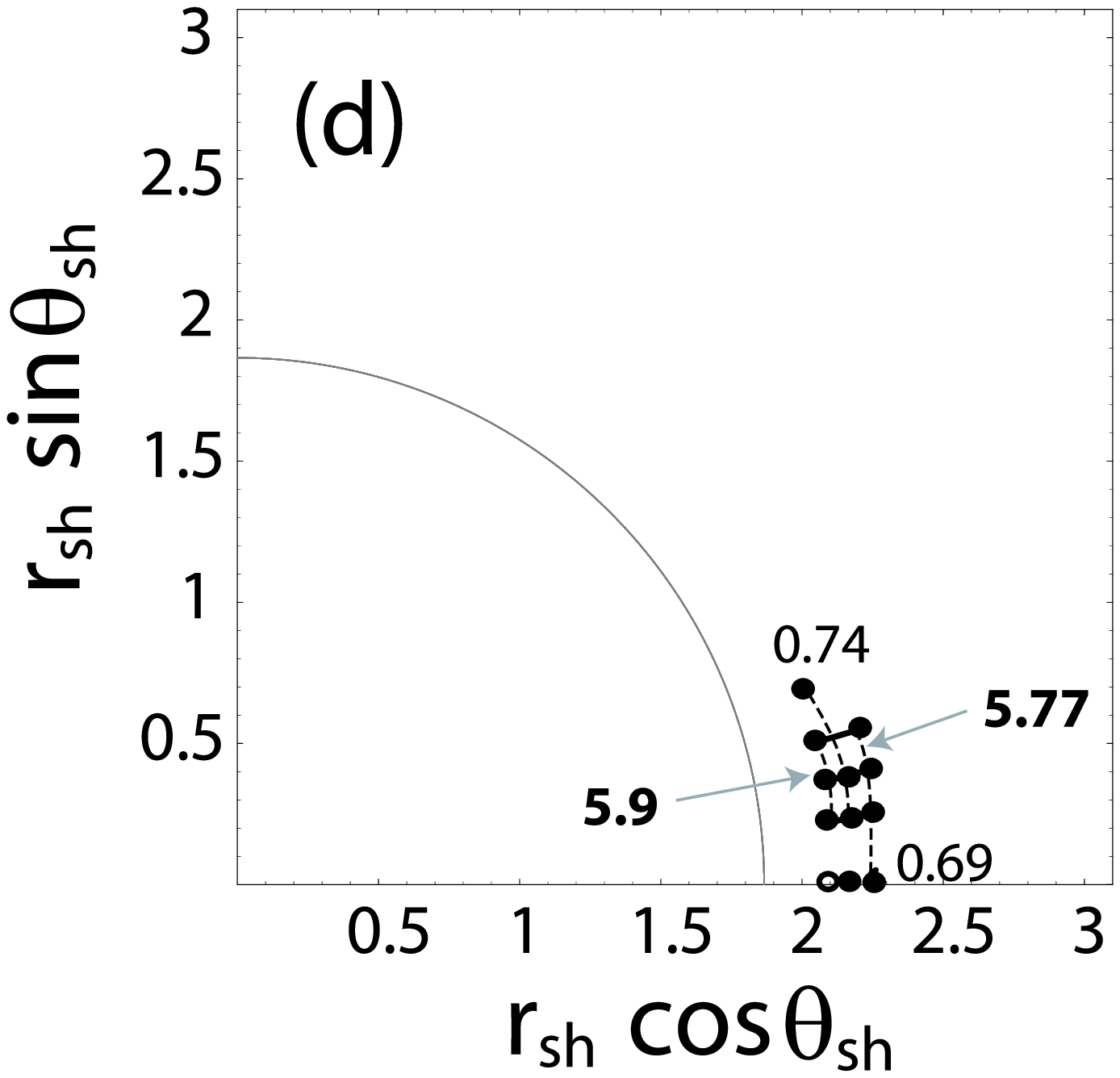}
    \includegraphics[angle=0, width=2.5in]{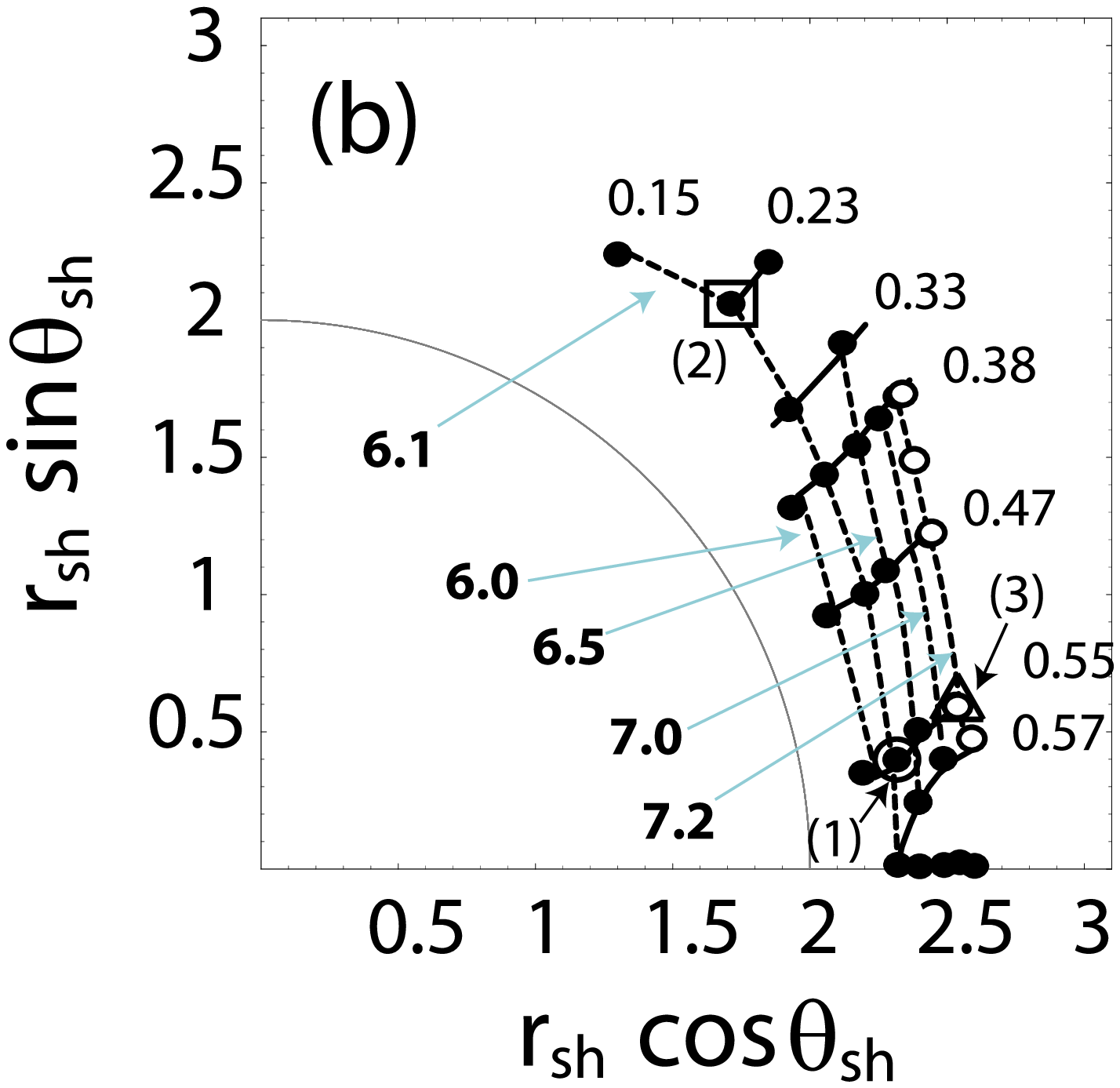}
    \includegraphics[angle=0, width=2.5in]{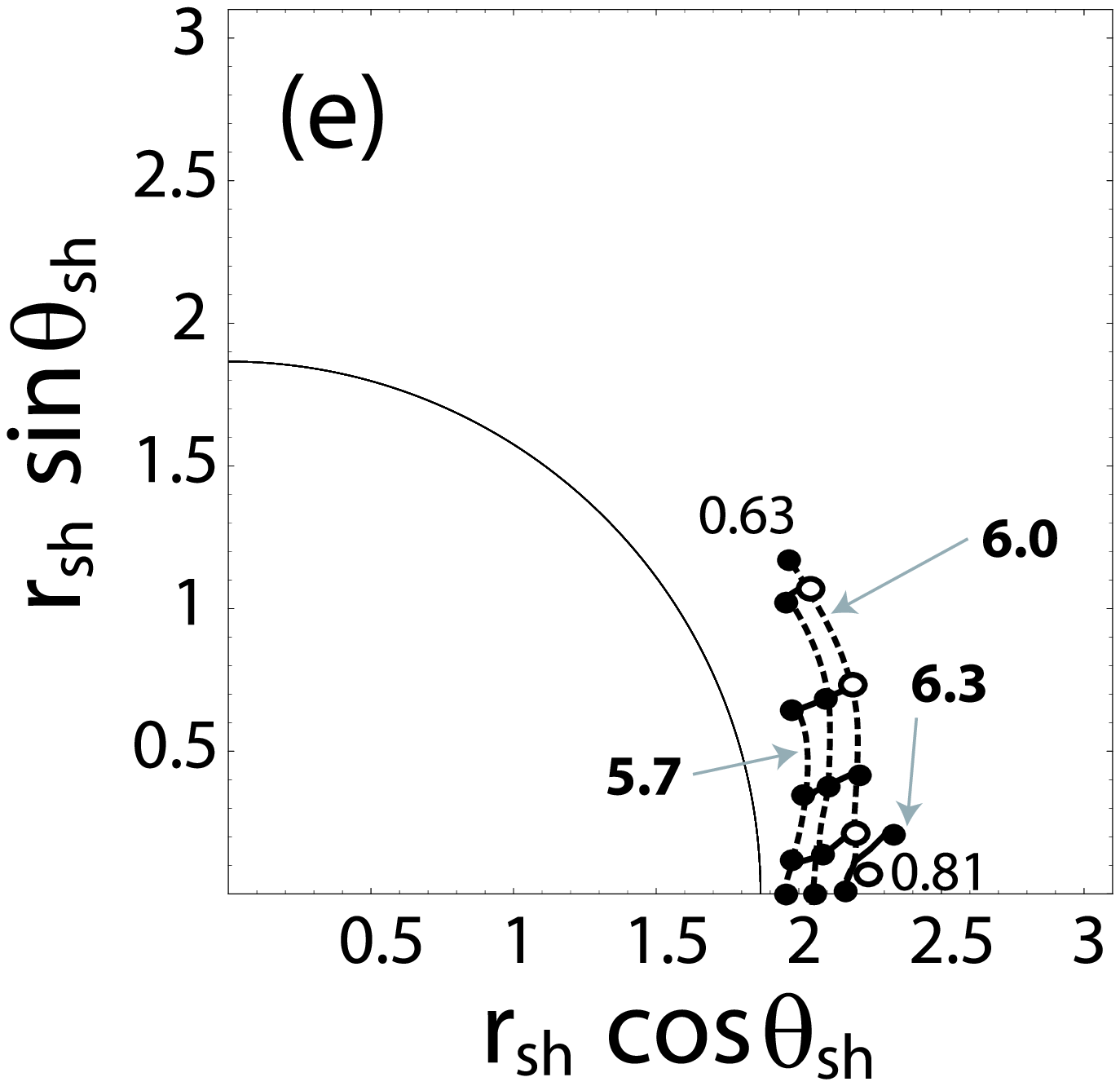}
    \includegraphics[angle=0, width=2.5in]{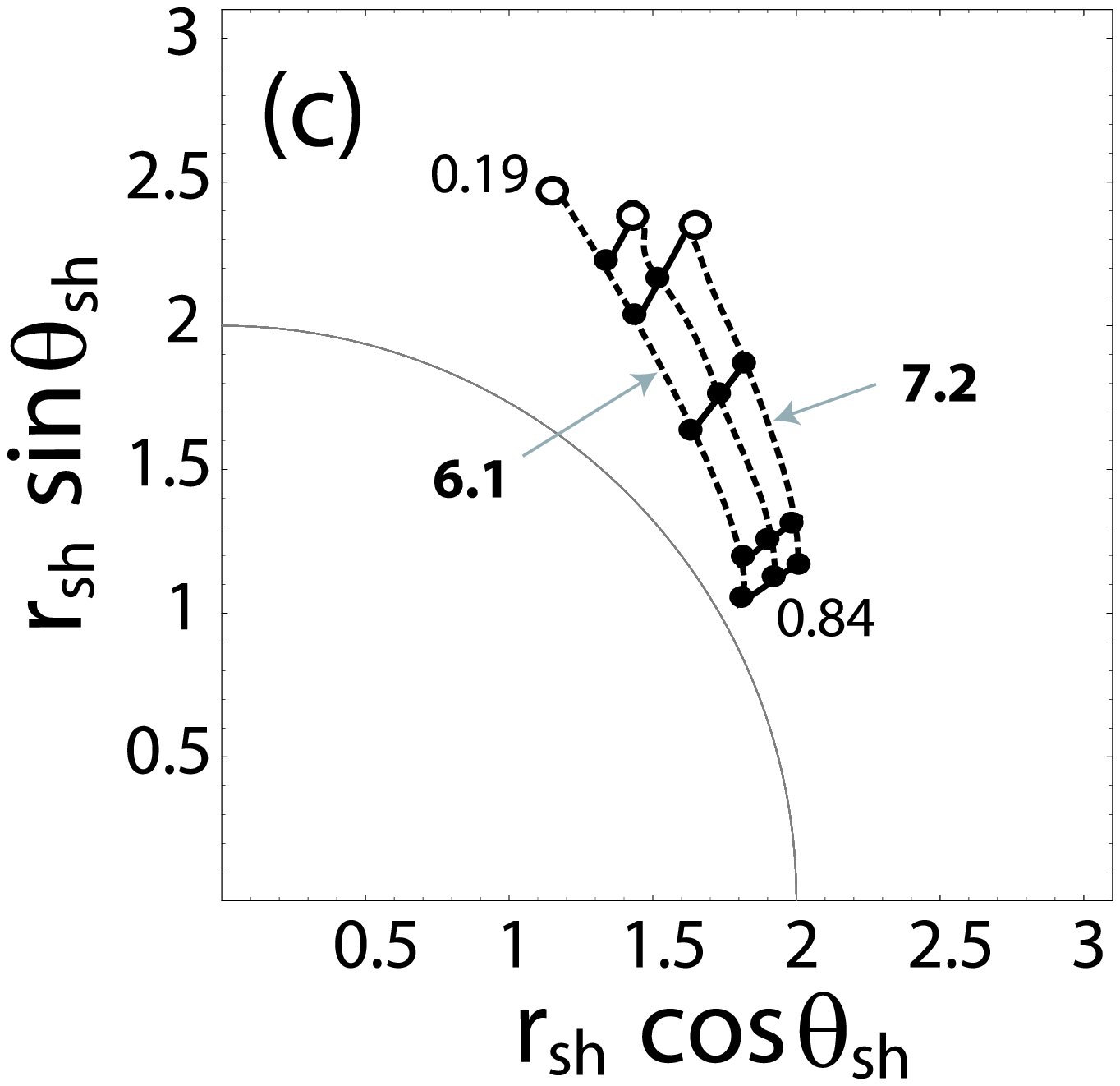}
    \includegraphics[angle=0, width=2.5in]{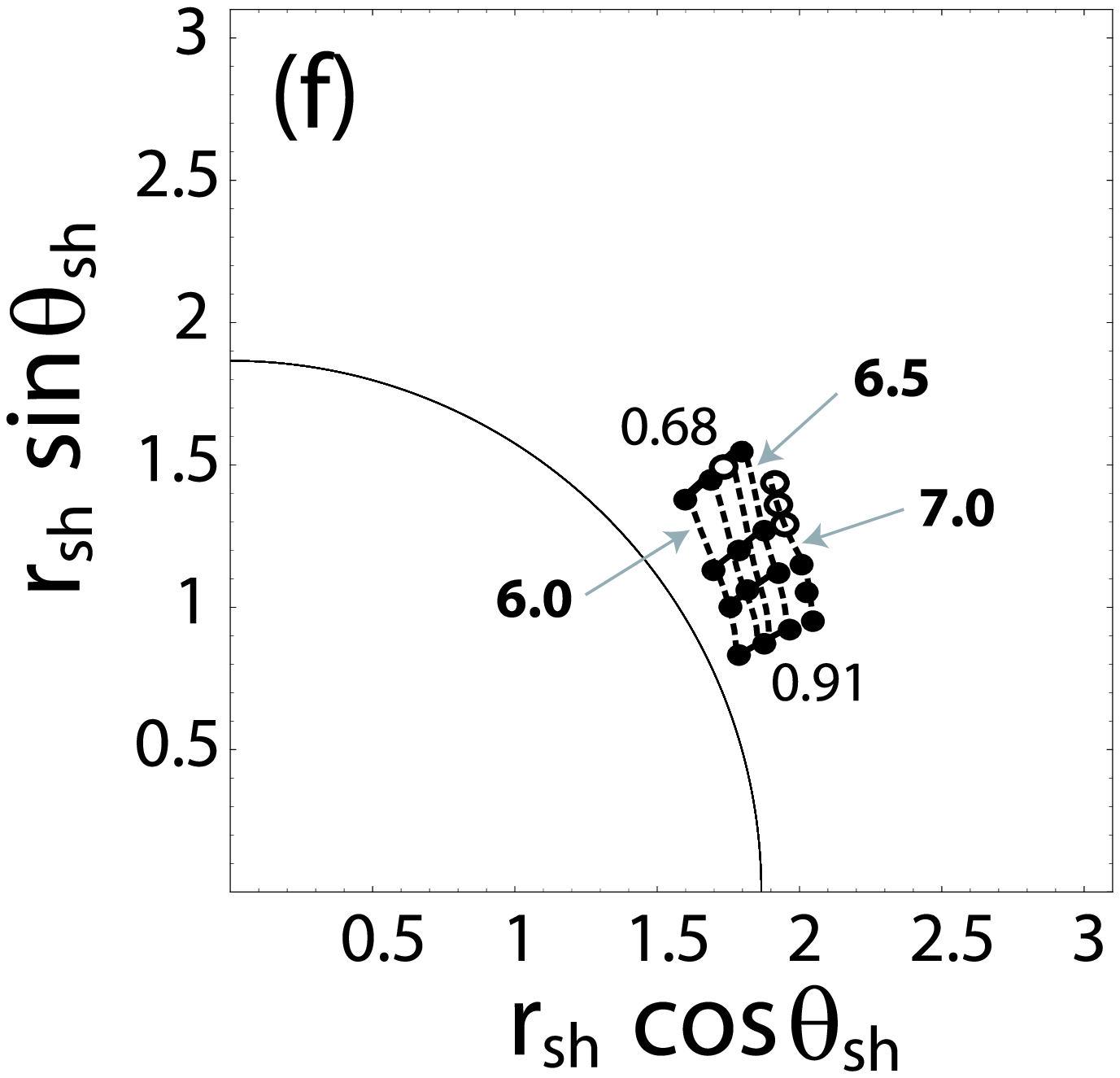}
    \caption{Acceptable parameter space for possible MHD shock formation on the $r_{\rm sh}$-$\theta_{\rm sh}$
    plane spanned by parameters ($\hat{E},\tilde{L} \Omega_F$). The filled circles ($\bullet$)
    represent the fast MHD shocks while the
    open circles ($\circ$) denote the intermediate MHD shocks.
    We choose (a) $\Omega_F=0.1$, (b) $0.14$ and (c)
    $0.2$ in the {\it left column} for $a=0$ and $\hat{\eta}=0.006$,
    while (d) $\Omega_F=0.21$, (e) $0.22$ and (f) $0.25$ in the {\it right column} for $a=0.5$ and $\hat{\eta}=0.005$.
    Solid curves
    labelled by the thin numbers denote constant
    $\tilde{L} \Omega_F$-curves,
    while dashed curves labelled by
    bold numbers denote constant $\hat{E}$-curves. The labels (1)-(3) in (b)
    denote the selected models. See Table~\ref{tab:tbl-1} for the
    characteristic parameters for these models.
    } \label{fig:fig2}
\end{figure}

We first show in Figure~\ref{fig:fig2} the possible shock region in
the poloidal plane ($r_{\rm sh},\theta_{\rm sh}$) for different
($\hat{E},\tilde{L} \Omega_F$) sets, where we can uniquely obtain a
point on the poloidal plane ($r_{\rm sh}, \theta_{\rm sh}$) for a
set of $(\hat{E},\tilde{L} \Omega_F)$. (a) $\Omega_F=0.1$, (b)
$0.14$, and (c) $0.2$ for the $a=0$ case and (d) $0.21$, (e) $0.22$
and (f) $0.25$ for the $a=0.5$ case. Along each curve in these
figures, either $\hat{E}$ or $\tilde{L} \Omega_F$ is fixed while the
other parameter (either $\tilde{L} \Omega_F$ or $\hat{E}$) is
varied, as indicated by the numbers. For a more detailed discussion
of the global nature of the trans-magnetosonic accretion solutions,
we select some representative shock solutions. The labels (1)-(3)
denote the selected models: the model (1) near-equatorial shock, the
model (2) mid-latitude shock with the same energy as (1), and the
model (3) near-equatorial shock with larger energy. See
Table~\ref{tab:tbl-1} for the characteristic parameters for these
models. As mentioned in \S2 the shock location $r_{\rm sh}$ is
constrained by the shock condition in
equation~(\ref{eq:shock-condition}) and the trans-magnetosonic
property.

%\textbf{We should mention that in the actual parameter search
%procedure, we find that $\eta$-dependence of the shock formation
%is very weak compared with the other parameters. So, in
%Figure~\ref{fig:fig2}, we show the possible shock
%parameter space by $\hat{E}$ instead of $\hat{\eta} \hat{E}$, and
%we use a certain specified constant $\hat{\eta}$-value to plot the
%diagram.}

Here, we should mention that, in the black hole magnetosphere with
a given black hole spin $a$ and magnetic field distribution
$\delta$, we can obtain the {\it cubic} parameter space ($\hat{E},
\tilde{L} \Omega_F$ and $\Omega_F$),
%---[ cut by MT ]
% ; note that the parameters $a$ and $\delta$ are fixed for a given
% black hole magnetosphere model,
%---//
and the dependence of various shock solutions [e.g.,
$\lambda(\theta_{\rm sh})$] on $\hat{\eta}$ is not focused here
since we find that it is very weak compared with the other
parameters. This is because $\hat{\eta} \hat{E} \sim$ constant once
$\tilde{L}$ and $\Omega_F$ are specified. That is, larger (smaller)
$\hat{\eta}$ and smaller (larger) $\hat{E}$ yields roughly the same
shock solutions for a fixed $\tilde{L} \Omega_F$ (see Paper I for
details). We thus show the possible shock parameter space by
$\hat{E}$ instead of $\hat{\eta} \hat{E}$ and use a certain
specified constant $\hat{\eta}$-value to plot the diagram.
Figure~\ref{fig:fig2} is a map of this cubic parameter space onto
the ($r_{\rm sh}, \theta_{\rm sh}$) plane, corresponding to a slice
of such a cubic parameter space by the $\Omega_F=$constant planes.
Thus, under the considered magnetosphere, physically valid MHD
shocks are restricted to this cubic space.

As $\hat{E}$ increases, the shock location $r_{\rm sh}$ tends to
shift further out for a given specific angular momentum
$\tilde{L}$. This is because the upstream plasma with very large
$\hat{E}$ generally means large kinetic energy, too, making it
easier for the fluid speed to exceed the magnetosonic speed.
Similar results are obtained in the HD shock formation (see, Lu \&
Yuan 1998 for the equatorial shocks and FT 2004 for the
non-equatorial shocks). Then, the shock can form sooner (i.e.,
further out). In order for accretion to effectively operate near
the mid to high latitudes, the specific angular momentum
$\tilde{L}$ needs to decrease to reduce the centrifugal force. The
accretion cannot take place otherwise. We have confirmed in
separate calculations that larger $\Omega_F$ can allow the MHD
shocks to form closer to the rotation axis with $\theta_{\rm sh}$
as small as $\sim 5\degr$.

\begin{figure}[t]% ------------------------------------- Figure~3
    \centering
    \epsscale{0.5}
    \includegraphics[angle=0, width=2.5in]{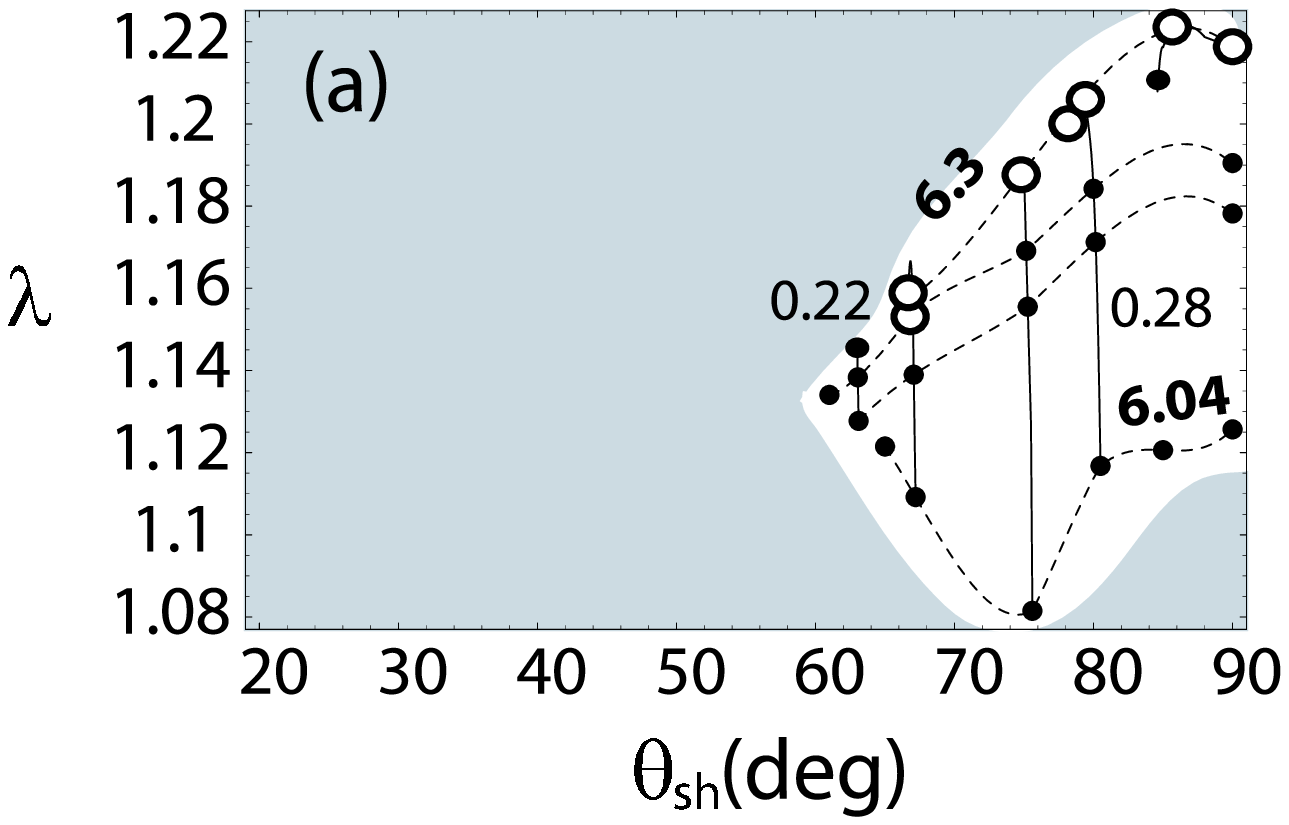}\includegraphics[angle=0, width=2.5in]{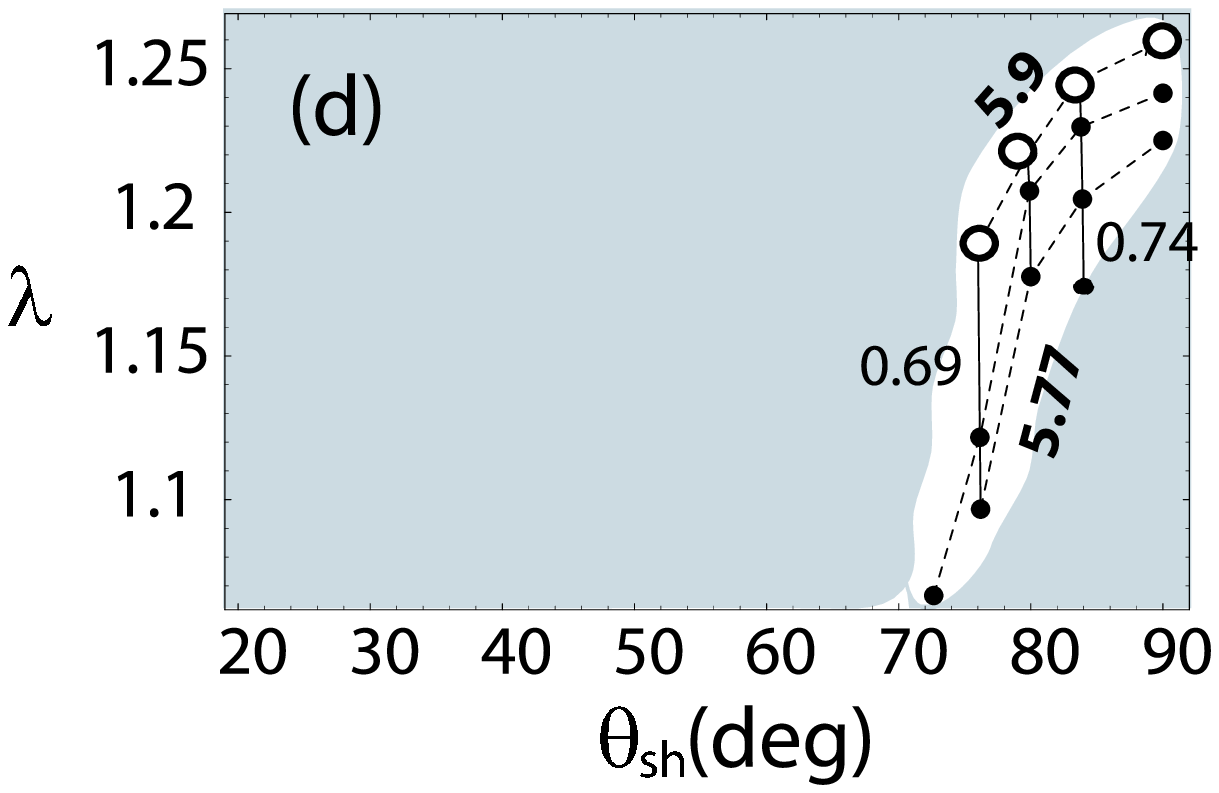}
    \includegraphics[angle=0, width=2.5in]{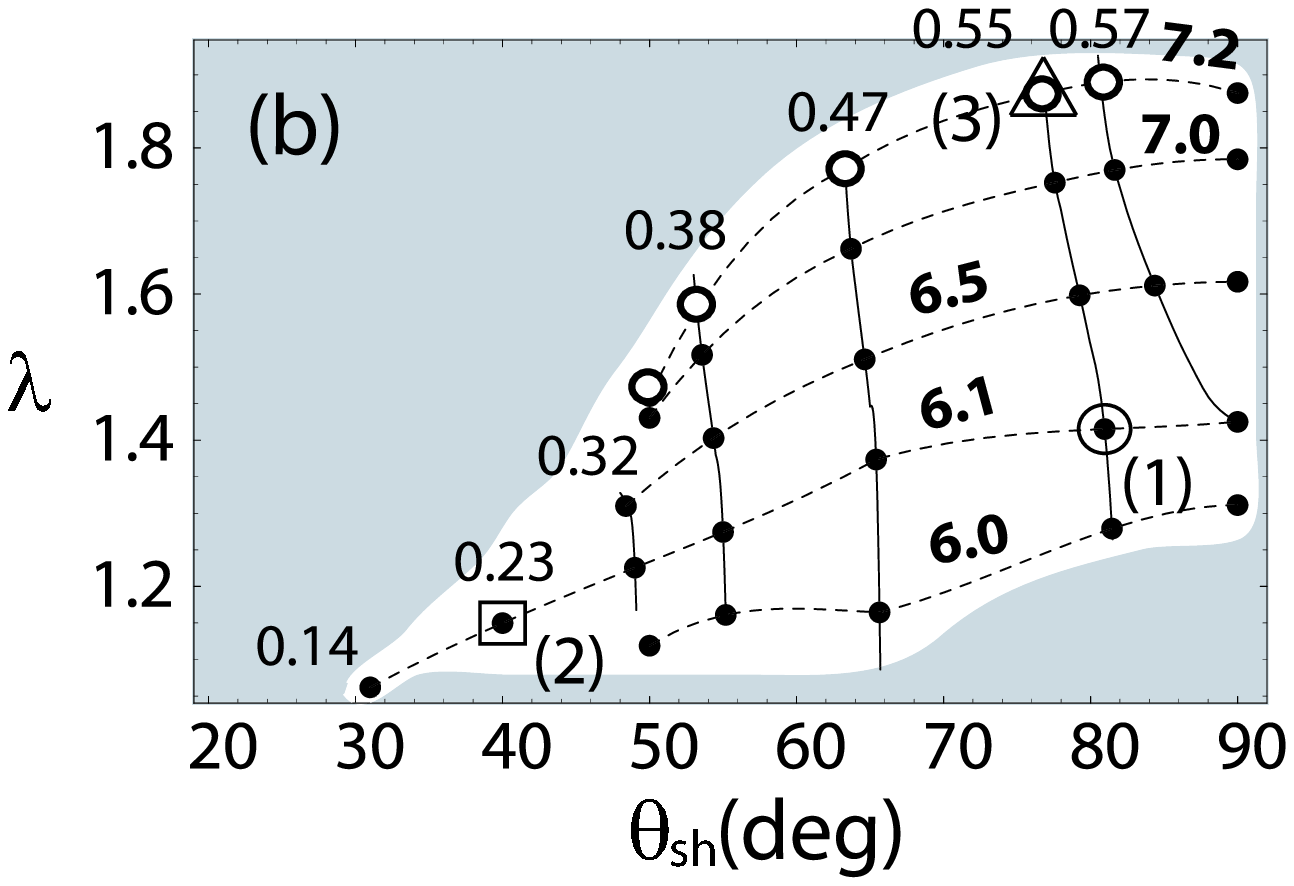}\includegraphics[angle=0, width=2.5in]{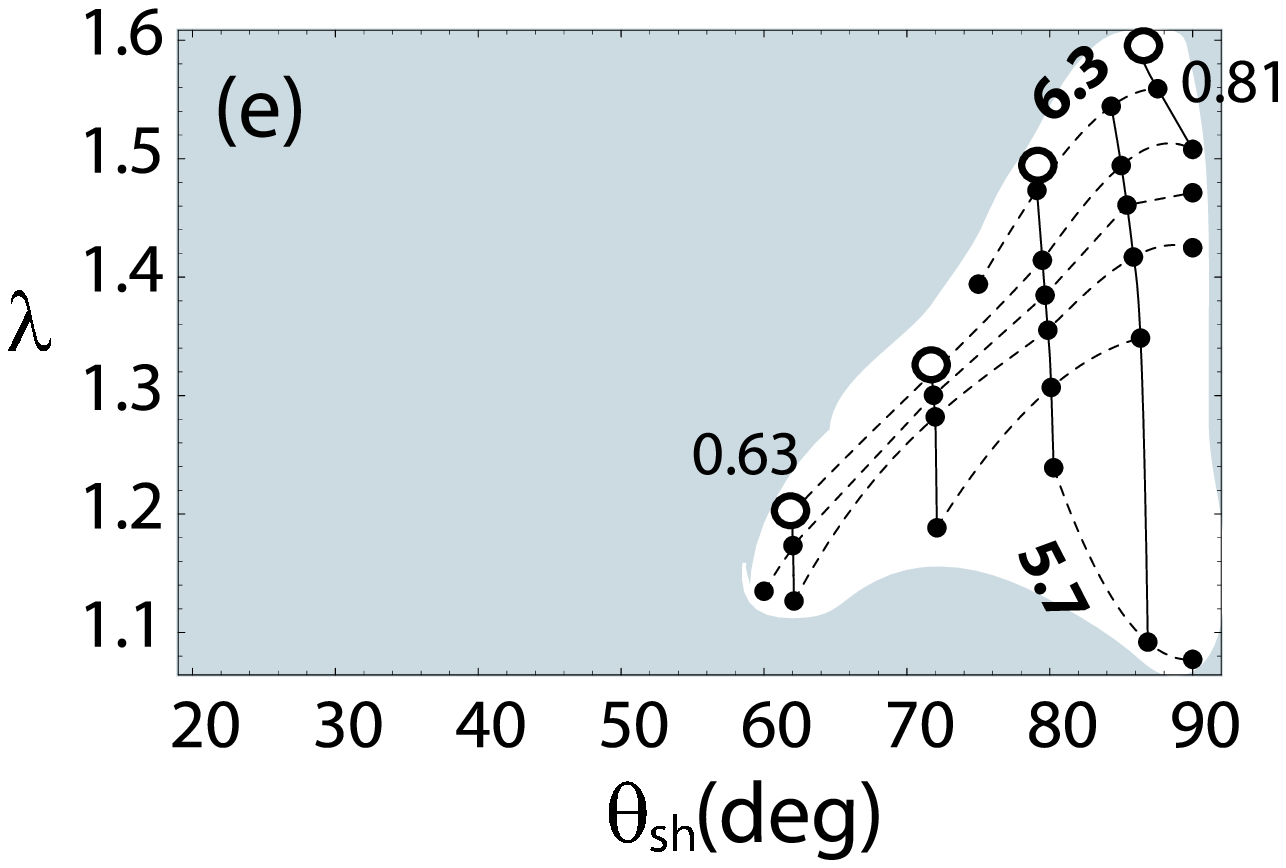}
    \includegraphics[angle=0, width=2.5in]{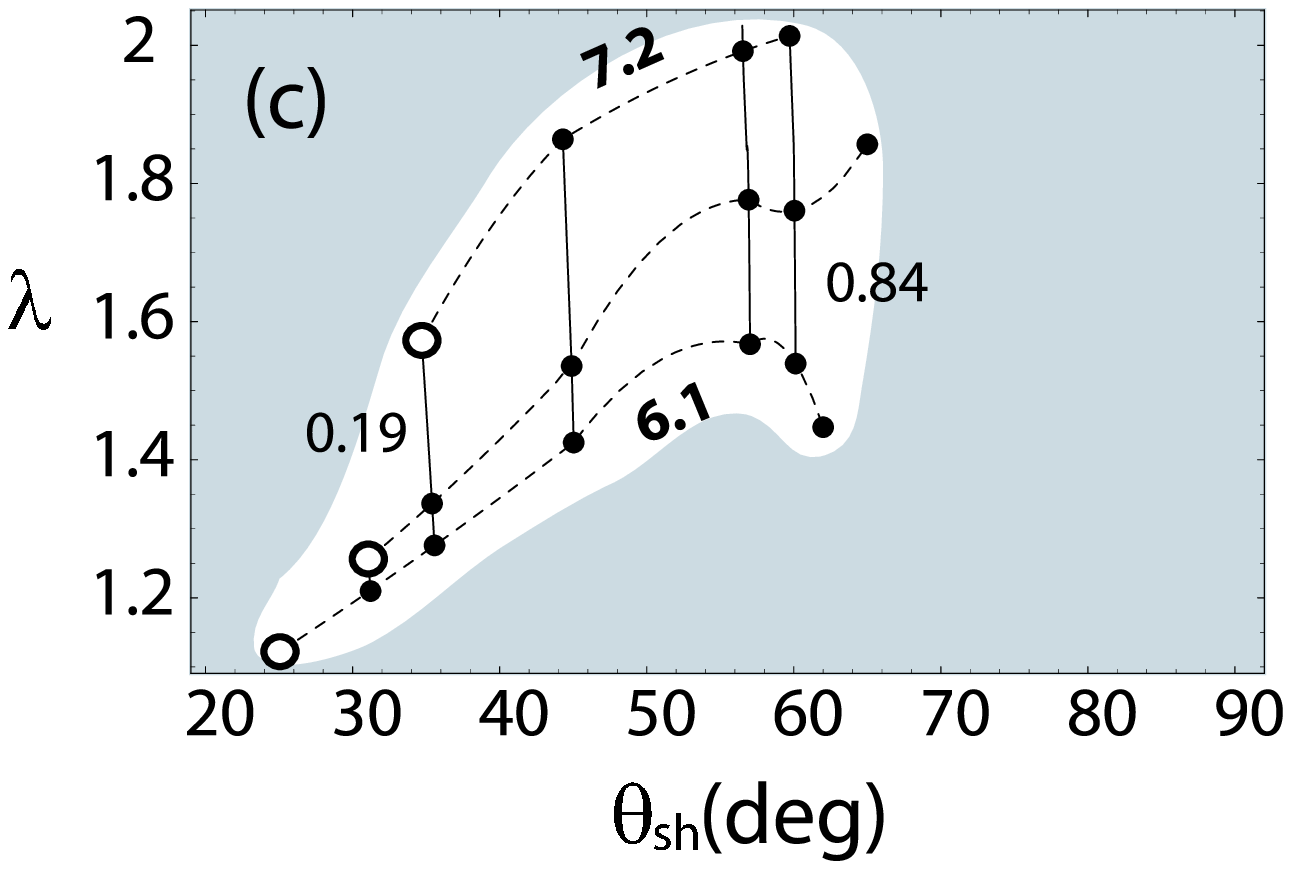}\includegraphics[angle=0, width=2.5in]{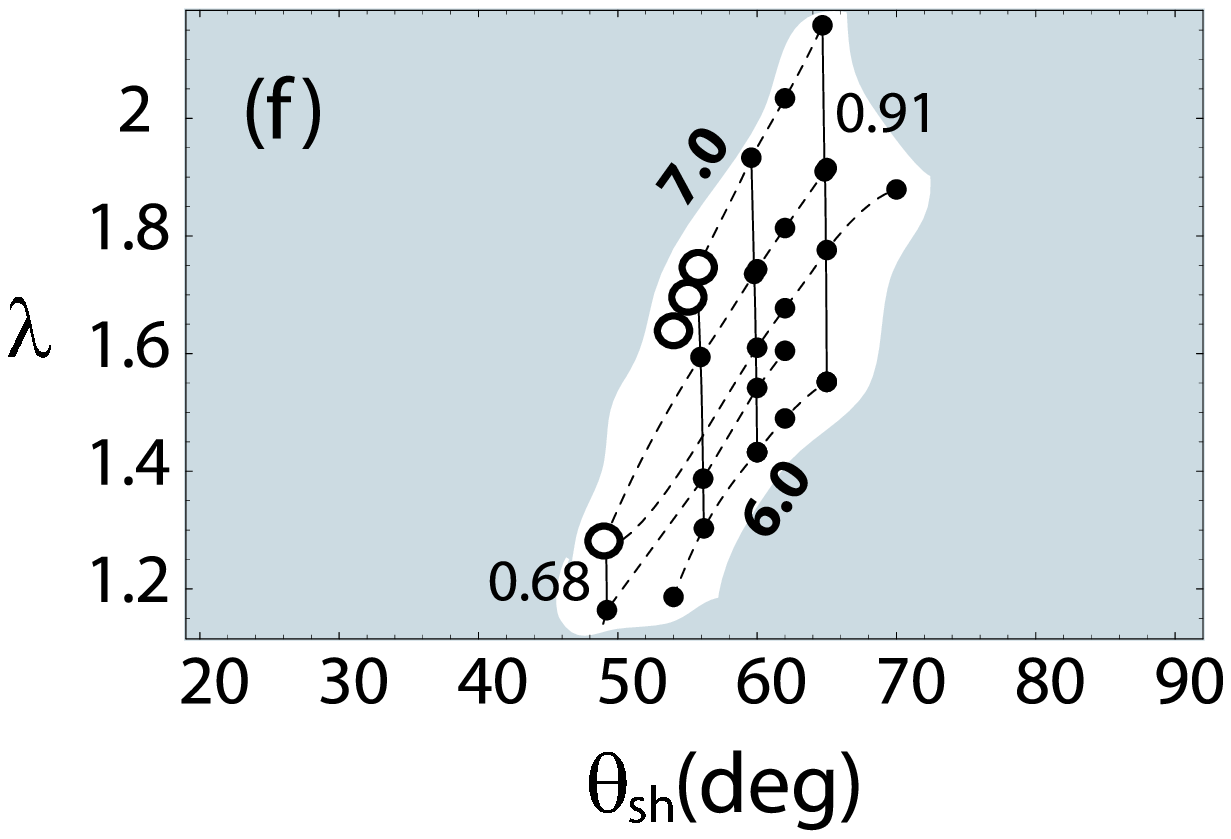}
    \caption{Shock compression ratio $\lambda$ as a function of polar angle $\theta_{\rm sh}$
    for various $\hat{E}$
    and $\tilde{L} \Omega_F$ corresponding to the solutions in Figure~\ref{fig:fig2}.
    The shaded regions
    represent the regions where no shock-included
    trans-magnetosonic solutions are possible (forbidden regions)
    for any $(\hat{E}, \tilde{L} \Omega_F)$ combination for a fixed $(\Omega_F,\hat{\eta})$.
    The other notations are the same as in Figure~\ref{fig:fig2}.} \label{fig:fig3}
\end{figure}

Next, let us examine the parameter space that allows the shock
formation. In Figure~\ref{fig:fig3} we display (fast and
intermediate) MHD shock solutions for various energy $\hat{E}$,
specific angular momentum, $\tilde{L}$ and polar angle $\theta_{\rm
sh}$ for fixed angular velocity of the field line $\Omega_F$. It is
first noted that the allowed shock region is continuous but limited
in the parameter space. For a given set of $(\hat{E},\tilde{L}
\Omega_F)$ the shock strength $\lambda$ generally becomes stronger
as $\theta_{\rm sh}$ increases (i.e., the closer to the equator, the
stronger the shocks). This is also seen in the non-equatorial HD
shocks (see FT04). The obtained qualitative pattern appears to be
independent of the choice of $\Omega_F$ and the black hole spin $a$
(compare the left and right columns in Figure~\ref{fig:fig3}). We
find that there exists the allowed region for the shock formation in
the parameter space spanned by ($\hat{E},\tilde{L} \Omega_F$) for a
fixed $\Omega_F$ and a given magnetosphere model. Outside the
allowed region (shaded region), physically valid shock solutions
disappear for the two reasons: (i) the shock condition by
equation~(\ref{eq:shock-condition}) is no longer met as $\hat{E}$
decreases (see, e.g., Paper I) and (ii) the inner fast magnetosonic
point disappears as $\hat{E}$ increases. That is, as you slice the
entire parameter space along a constant $\tilde{L} \Omega_F$-curve
as in Figure~\ref{fig:fig3}, there exists a definite border line
between the shock allowed and forbidden regions determined by either
condition (i) or (ii) depending on the $\tilde{E}$-value.
Additionally, there is a constraint on the parameter space in terms
of the existence of the \Alfven~point. Because the parameter space
allowed by the requirement of the \Alfven~point is wider than those
by the conditions (i) and (ii), the allowed shock region shown in
Figure~\ref{fig:fig3} satisfies all the conditions by (i), (ii) and
the \Alfven~point.

In general we have confirmed that the minimum value of $\theta_{\rm
sh}$ is determined by the above condition (i) while the maximum
value of $\theta_{\rm sh}$ by the condition (ii). For accretion to
take place in higher latitude regions (i.e., smaller $\theta_{\rm
sh}$), the angular velocity of the field line $\Omega_F$ is forced
to increase. Because the range of $\tilde{L} \Omega_F$-value is
restricted by the existence of the \Alfven~points on the
shock-included trans-magnetosonic accretion solution (see Paper I
for details), $\Omega_F$ must increase as $\tilde{L}$ decreases in
higher latitude regions for a shock to occur, where $\tilde{L}
\propto \sin^2\theta_{\rm sh}$. Thus, the allowed region in
Figure~\ref{fig:fig3} topologically shifts as $\Omega_F$ changes.
That is, the shock is allowed more in the mid-high latitudes as
$\Omega_F$ increases. In other words, high latitude shock formation
can only be allowed by large $\Omega_F$. In this case, the shock
tends to become stronger with increasing $\Omega_F$. This feature,
due to the rotation of the magnetic field line, is a new aspect
inherent to the MHD shocks which would not exist in the HD shock
properties.
%---[ cut by MT ]
% On the other hand, specific
% angular momentum $\tilde{L}$ needs to be smaller near high
% latitude regions for accretion to take place.
%---//
We have obtained diagrams similar to Figure~\ref{fig:fig3} for
different $\hat{\eta}$-values and find that the
$\hat{\eta}$-dependence of the shock solutions is very weak. As to
the dependence on energy $\hat{E}$, it is apparent that the shock
strength $\lambda$, for instance, is very sensitive to the change in
$\hat{E}$ [see, e.g., the models (1) and (3) along the solid curve].
On the contrary, changing $\hat{E}$ under the fixed $\tilde{L}
\Omega_F$ value has only a weak effect on shifting the polar angle
$\theta_{\rm sh}$. For a given $\hat{E}$, the shock can occur over a
wide range of angle $\theta_{\rm sh}$ [see, e.g., the models (1) and
(2) along the dashed curve].

The topological nature of the parameter space for $a=0.5$ is
essentially the same as for the $a=0$ case (compare the left and
right columns in Figure~\ref{fig:fig3}), in the sense that larger
$\Omega_F$ allows stronger shocks near the higher latitude regions.
However, for rotating black holes the shock region does not reach as
high latitudes as for non-rotating ones. We have also found that the
black hole spin $a$ generally amplifies the shock strength as in the
HD shocks. However, the coupling among various parameters in the
presence of the magnetic field complicates the situation. Around a
rotating black hole, it is seen that the allowed region in the polar
direction appears to be narrower compared with the $a=0$ case
(compare the left and right columns in Figure~\ref{fig:fig3}). This
is because there is a tighter constraint on the value for $\tilde{L}
\Omega_F$ (not for $\eta E$) that allows shock formation. This is
explained by considering the presence of the \Alfven~point on the
accretion solution. Because the location of the \Alfven~point is
determined by $\tilde{L} \Omega_F$, as $a$ changes, the allowed
range of $\tilde{L} \Omega_F$ is varied (narrowed down) for the
\Alfven~point to exist on the accretion solution \citep{TNTT90}.
This black hole spin effect for the \Alfven~point location is seen
in Figure~\ref{fig:fig3} for the allowed shock location.

We present other various shock properties for the postshock flows in
Figure~\ref{fig:fig4}; (a) shocked plasma temperature $\Theta$, (b)
postshock entropy-related accretion rate $\dot{\cal{M}}_2$ and (c)
postshock magnetization $\sigma_2$. The shock solutions marked by
(1)-(3) in Figure~\ref{fig:fig4} correspond to the models (1)-(3) in
Figure~\ref{fig:fig2}b. Temperature $\Theta$ strongly depends on
$\hat{E}$. With increase of $\hat{E}$, the temperature can rise
significantly [e.g., compare the models (1) and (3)]. The
$\dot{\cal{M}}_2$ vs. $\theta_{\rm sh}$ behavior is roughly similar
to the $\Theta$ vs. $\theta_{\rm sh}$ behavior, except that there is
a peak angle in the mid-latitude. On the other hand, the downstream
magnetization $\sigma_2$ acts quite differently from the other two.
With small increase of $\hat{E}$, $\sigma_2$ decreases significantly
[compare the models (1) and (3)]. That is, $\Theta$ and $\sigma_2$
show a clear anti-correlation, as expected from the conserved energy
equation (\ref{eq:E}). We also see a trend that the equatorial
shocked plasmas are generally more magnetized than the
non-equatorial ones. In separate calculations, we have confirmed
that increasing $\Omega_F$ will amplify the downstream magnetization
$\sigma_2$ while it will reduce temperature $\Theta$. Furthermore,
in Figure~\ref{fig:fig4}c for a fixed $\hat{E}$ the downstream
magnetization $\sigma_2$ becomes larger as the $\tilde{L}
\Omega_F$-value becomes larger. Therefore, this is a good indicator
for estimating the degree of magnetization of the shocked plasma.

In \S2.2 we assumed conical magnetic fields, which would be
reasonable around a black hole. However, the magnetospheric
configuration other than the $\delta=0$ case may be possible, where
$\delta$ parameterizes the cross-section of the magnetic flux.
Therefore, we will here explore the effects caused by some other
magnetic field configurations. Figure~\ref{fig:fig5} illustrates the
dependence of shock strength $\lambda$ on $\delta$. The pure
split-monopole field geometry on the poloidal plane is realized with
$\delta=0$, while the outward divergent poloidal field is
parameterized by $\delta>0$. We find that the MHD shock tends to
become stronger as $\delta$ becomes larger. This is because for
$\delta>0$ the accreting plasma is more accumulated (or
concentrated) toward a smaller cross-sectional area with decreasing
$r$ due to the tapered magnetic field. With increasing $\delta$, the
accreting plasma in such a field geometry would also become slower.
Furthermore, the outward magnetic pressure would be more effective
in that case. A larger deviation from a purely radial field
configuration tends to not only allow the stronger fast shock
$\lambda$, but also enhance the shocked plasma temperature $\Theta$.
Note that the increase in $\delta$ shifts the inner fast
magnetosonic point inward, while the outer one remains almost the
same \citep{Takahashi02}. The valid shock location must lie between
these outer and inner fast magnetosonic points. Since the location
of the fast magnetosonic points shift as $\delta$ changes, the
allowed shock location also changes.

\subsection{Properties of the Shock-Included Trans-Magnetosonic Solutions}

In this section we consider the global, shock-included
trans-magnetosonic solutions for the selected cases [i.e., models
(1)-(3)]. We present in Figure~\ref{fig:fig6} several representative
solutions, where the models (1)-(3) are selected from
Figure~\ref{fig:fig2}b. That is, the model (1) near-equatorial
shock, the model (2) mid-latitude shock with the same energy as the
model (1) and the model (3) near-equatorial shock with greater
energy. These inflows all start from several Schwarzschild radii.

\begin{figure}[t]% ------------------------------------- Figure~4
    \centering
    \epsscale{0.5}
    \includegraphics[angle=0, width=2.4in]{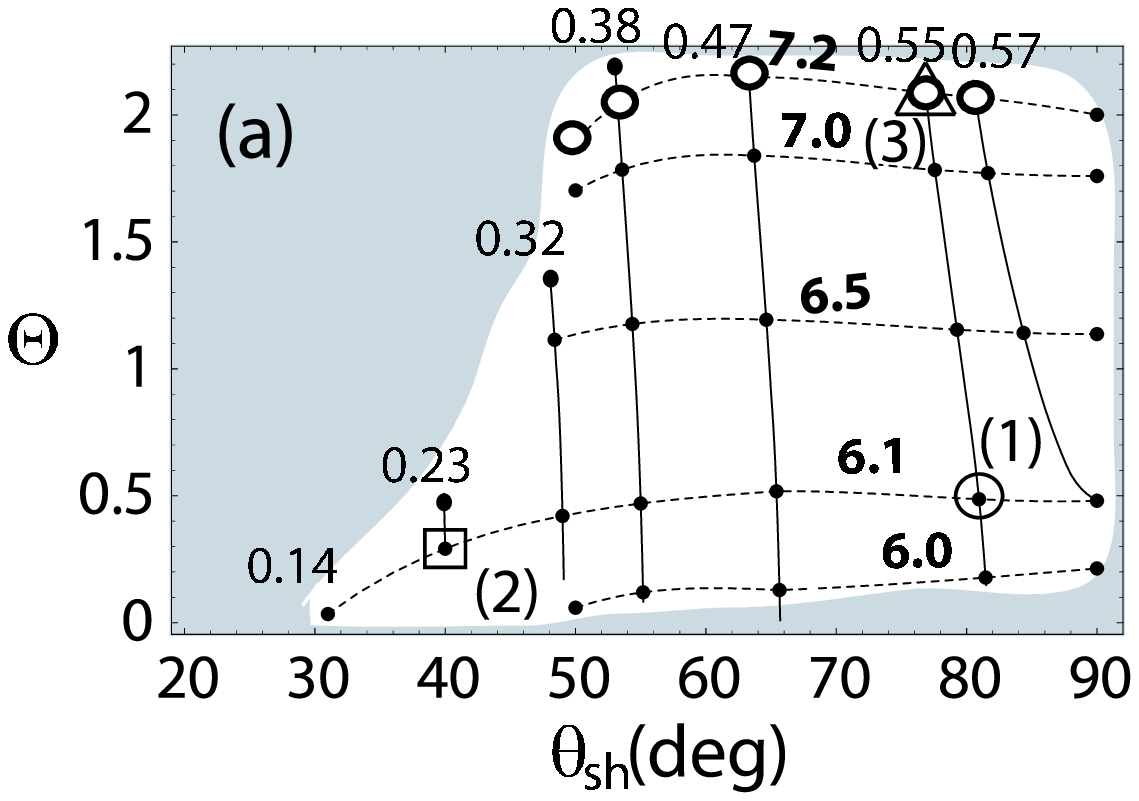}
    \includegraphics[angle=0, width=2.6in]{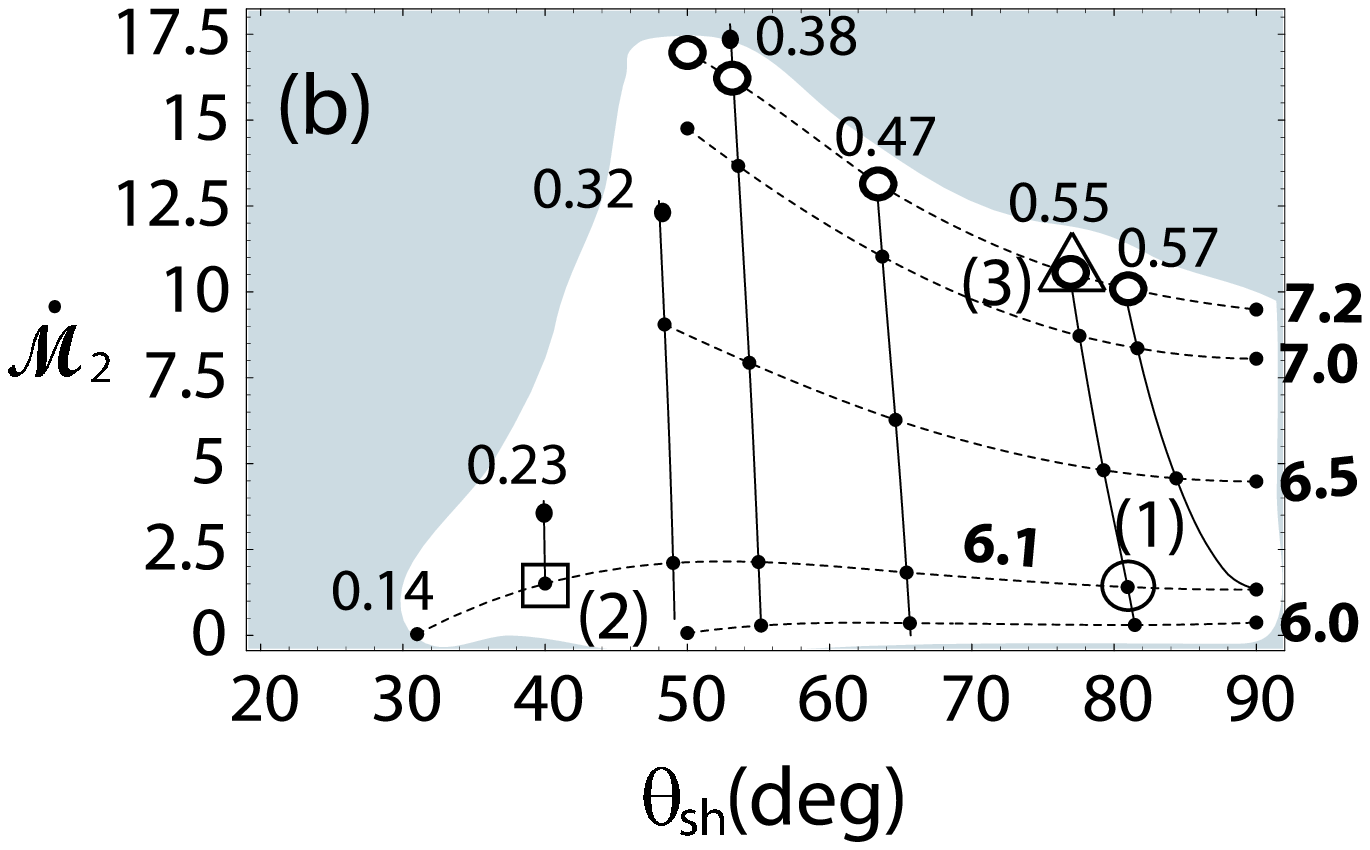}
    \includegraphics[angle=0, width=2.6in]{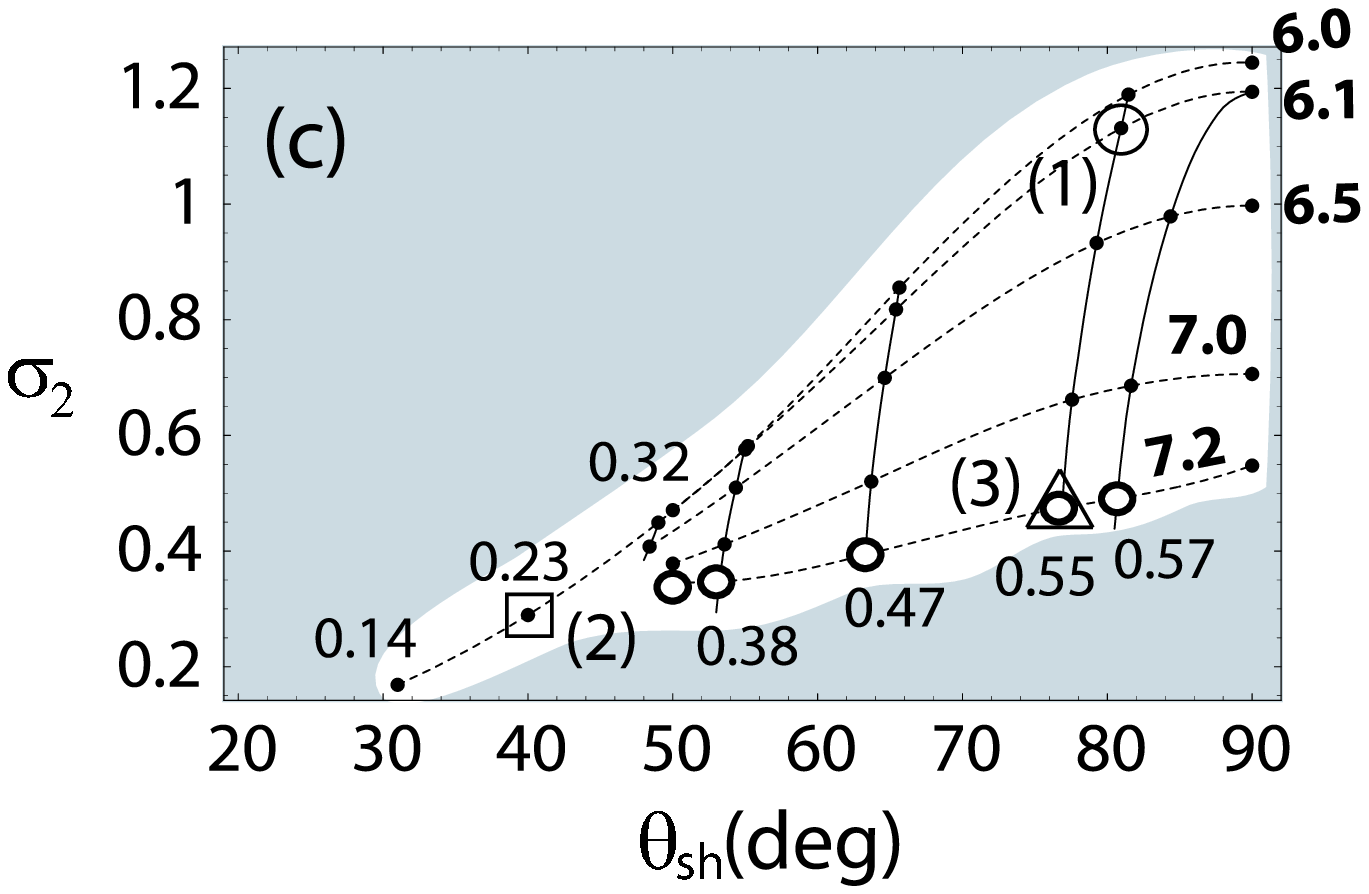}
    \caption{Other shock quantities as a function of $\theta_{\rm sh}$ corresponding to the solutions
    in Figure~\ref{fig:fig2}b: (a) Shocked plasma temperature $\Theta$, (b)
    postshock entropy-related accretion rate $\dot{\cal{M}}_2$,
    and (c) postshock magnetization parameter $\sigma_2$.
    The notations are the same as in Figure~\ref{fig:fig2}. }
    \label{fig:fig4}
\end{figure}

\begin{figure}[t]% ------------------------------------- Figure~5
    \centering
    \epsscale{0.5}
    \includegraphics[angle=0, width=2.5in]{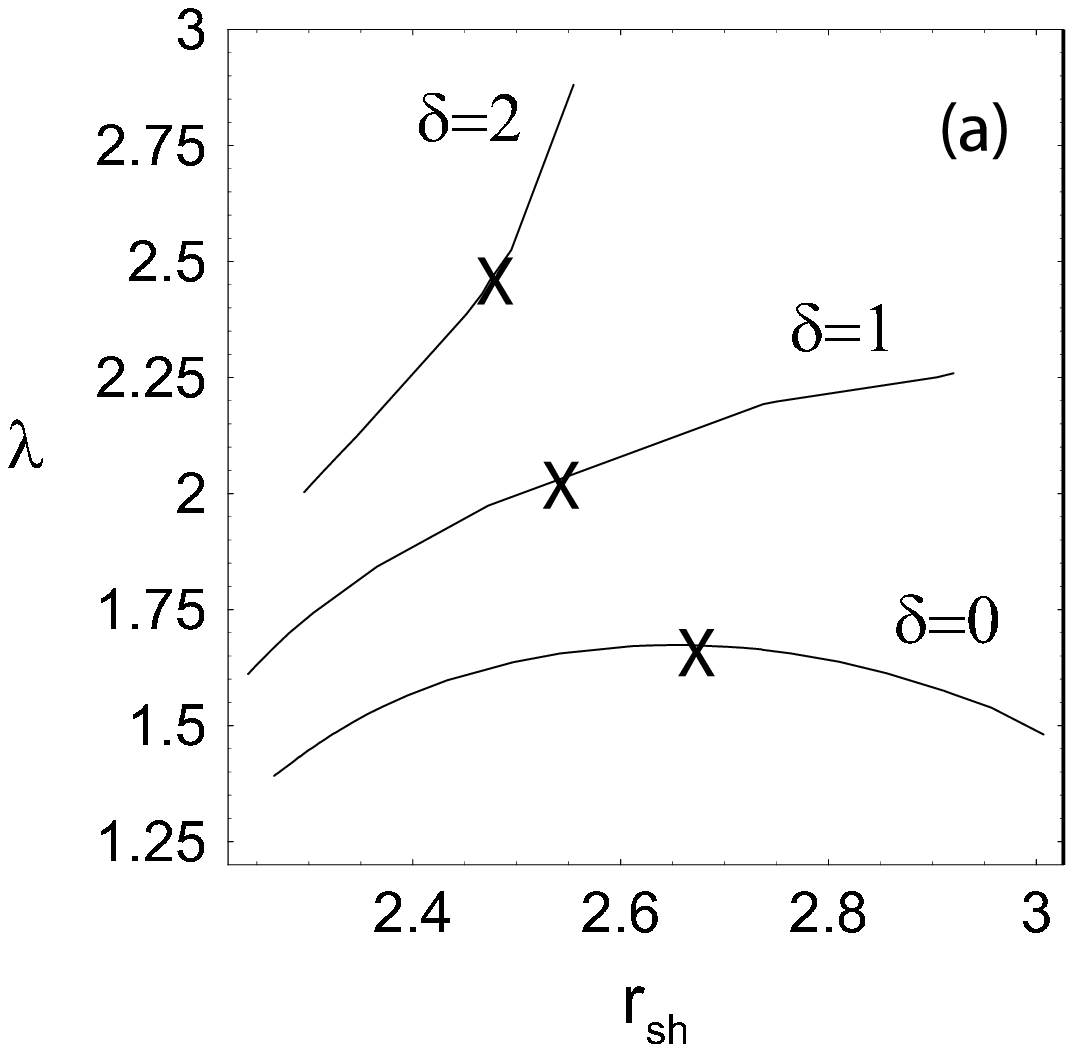}
    \includegraphics[angle=0, width=2.5in]{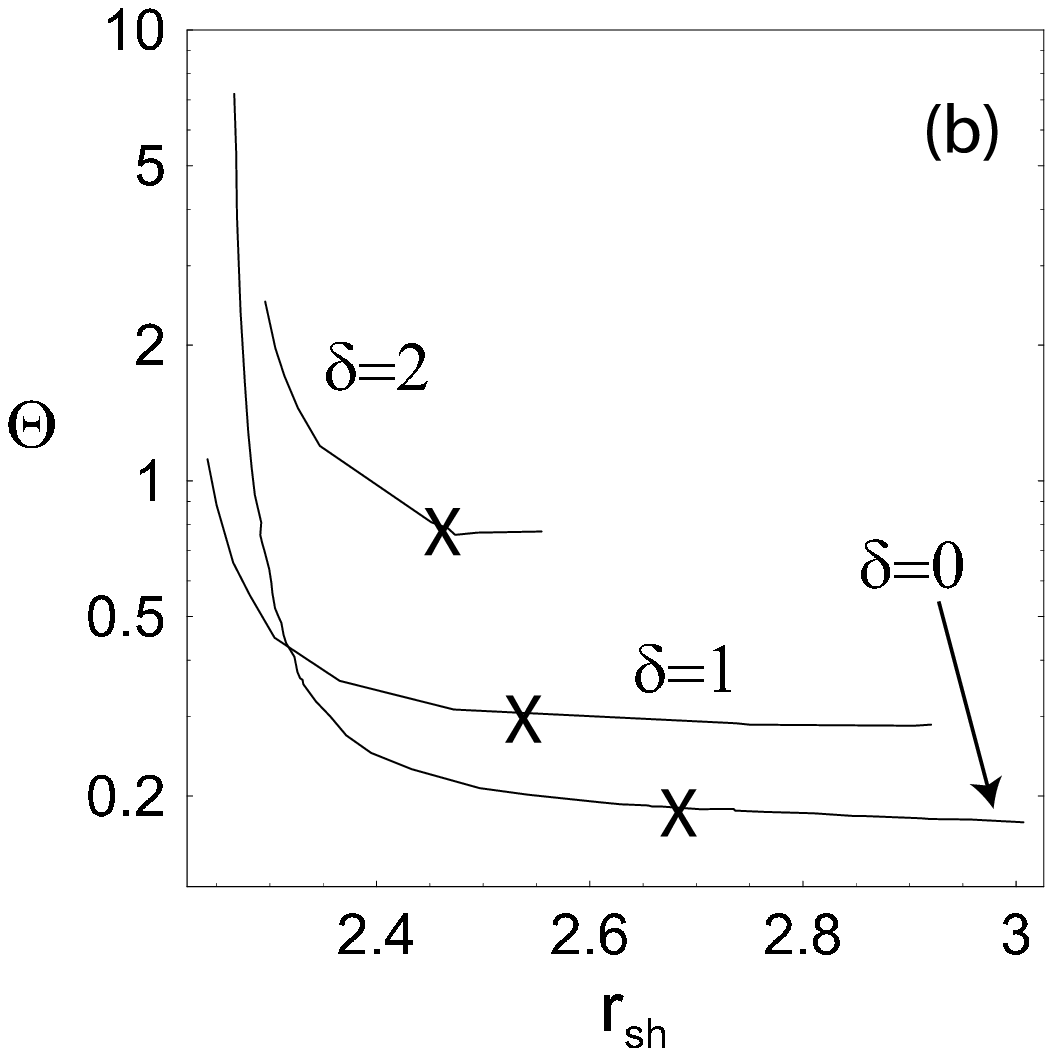}
    \caption{The dependence of the shock strength $\lambda(r_{\rm sh})$
    on $\delta$ for $\hat{E}=6.1,\tilde{L}=4.1$ and $a=0$.
    We set $\hat{\eta}=0.006$ for $\delta=0$, $\hat{\eta}=0.003$ for $\delta=1$ and $\hat{\eta}=0.001$
    for $\delta=2$.
    The crosses ($\times$) denote the anchor points. The intermediate MHD shocks are obtained to the right
    of the anchor points. }
\label{fig:fig5}
\end{figure}

\begin{figure}[t]% ------------------------------------- Figure~6
    \centering
    \epsscale{0.5}
    \includegraphics[angle=0, width=5.1in]{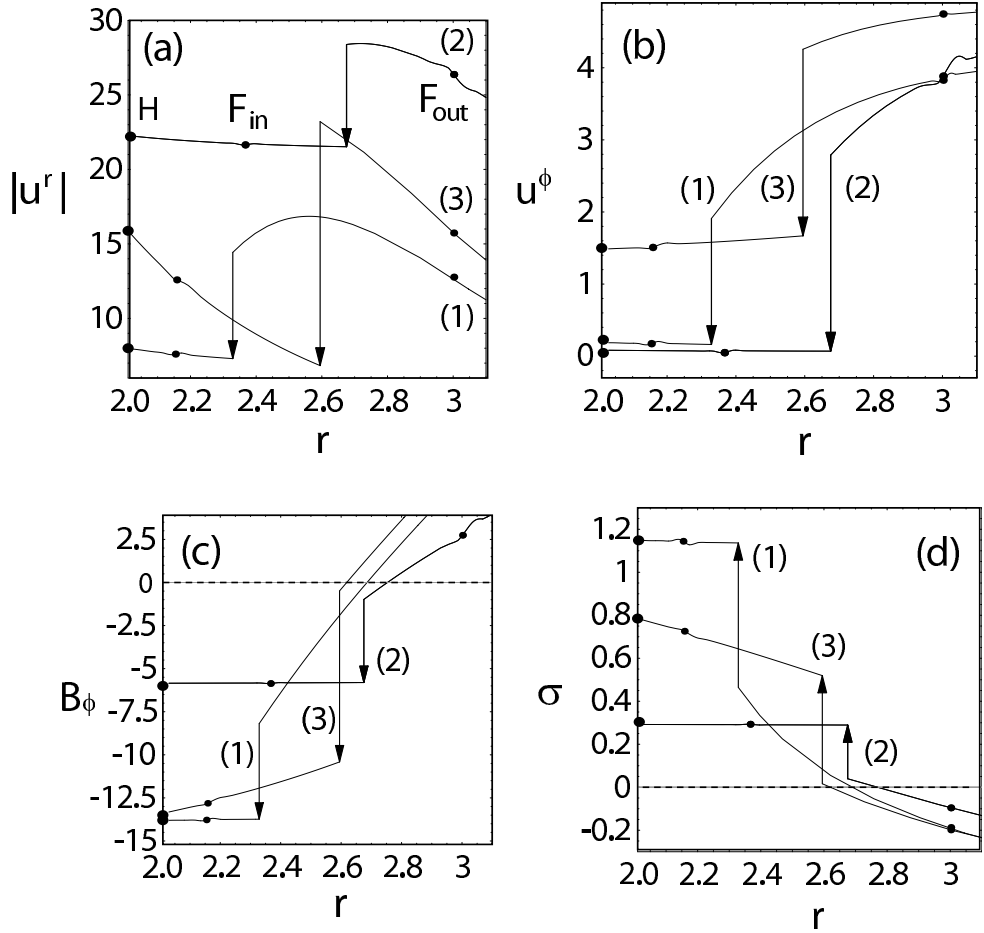}
    \caption{Shock-included trans-magnetosonic
    accretion solutions as a function of radius $r$: (a) the radial four-velocity $u^r$, (b)
    the toroidal four-velocity $u^\phi$, (c) the toroidal magnetic field
    $B_\phi$, and (d) the magnetization parameter $\sigma$.
    The MHD shock in the accretion solution ({\it vertical downward arrow}) is
    formed after the preshock solution passes through the outer fast magnetosonic
    point, F$_{\rm out}$. The postshock solution then passes through
    the inner fast magnetosonic point, F$_{\rm in}$, and falls onto
    the event horizon, H. The
    solutions labelled (1)-(3) correspond to the models (1)-(3) in Figure~\ref{fig:fig2}b.
    See Table~\ref{tab:tbl-1} for the model parameters. }
\label{fig:fig6}
\end{figure}

The model (1) shows a relatively small downstream kinetic motion
[i.e., $|u^r_2(r)|$ and $u^\phi_2(r)$] whose magnitude almost
remains the same until reaching the event horizon, meaning that the
external forces are balanced with the gravity (see
Figure~\ref{fig:fig6}a and b). For the total energy to be conserved,
a large fraction of the total energy then needs to be redistributed
to somewhere else. In the model (1), the major part of the energy is
redistributed to the magnetic field across the shock, and the
postshock magnetization $\sigma_2(r)$ is relatively large ($\sigma_2
\sim 1.1$ in Figure~\ref{fig:fig6}d), while the shocked plasma
temperature is not high ($\Theta \sim 0.5$ in
Figure~\ref{fig:fig4}a). Such an anti-correlation is expected from
the shock properties shown in Figure~\ref{fig:fig4}. Note that
$|B_{\phi1}(r_{\rm sh})| < |B_{\phi2}(r_{\rm sh})|$, as expected,
for the fast MHD shock.

The model (2) has the same energy as the model (1) except for a
smaller angular momentum $\tilde{L}$ at $\theta_{\rm sh}=40\degr$.
Since the toroidal magnetic field near the event horizon scales as
$B_\phi \sim B_{\phi,H} \propto \sin^2 \theta_H$ where the subscript
$H$ denotes the event horizon, smaller $|B_{\phi,2}|$ is required
for the model (2) as seen in Figure~\ref{fig:fig6}c. In this case,
the magnetic component of energy is accordingly smaller, and the
postshock magnetization $\sigma_2(r)$ behaves as shown in
Figure~\ref{fig:fig6}d. To compensate for a smaller magnetic energy
component, postshock kinetic energy becomes relatively larger (see
Figure~\ref{fig:fig6}a). As a result, heating is not significant
($\Theta \sim 0.3$ in Figure~\ref{fig:fig4}a). We also find that the
shock location $r_{\rm sh}$ shifts outward and in higher latitude
regions as in the HD case (see FT04). For example, as $\theta_{\rm
sh}$ changes from $81\degr$ in the model (1) to $40\degr$ in the
model (2), the shock location shifts outward from $r_{\rm sh}=2.3$
to $2.7$ (see Table~\ref{tab:tbl-1}). This is because the
centrifugal force barrier becomes stronger as $\theta_{\rm sh}$
decreases.

As we discussed in the previous sub-section, in
Figure~\ref{fig:fig6} the value of $\tilde{L} \Omega_F$ is larger in
the model (1) than that in the model (2) for a fixed $\hat{E}$. The
downstream magnetization parameter $\sigma_2(r)$ in
Figure~\ref{fig:fig6}d is thus greater in the model (1), meaning
that the model (1) shows the more magnetized downstream plasma.

For the model (3) we choose a larger energy, $\hat{E}=7.2$ to see
the energy dependence of the shock solution. For comparison, the
values for the angular momentum and the polar angle are roughly kept
the same as in the model (1). Despite the larger total energy, the
toroidal magnetic field strength at the event horizon $|B_{\phi,H}|$
has a similar value to that in the model (1) because of the boundary
condition at the event horizon. The energy at the shock front in
this case is then redistributed more to the thermal and kinetic
energies. In fact, we see that the downstream radial acceleration
$|d u^r_2(r)/ dr|$ is larger, and the downstream toroidal velocity
$u^\phi_2(r)$ is also larger, compared with the models (1) and (2).
As expected from Figure~\ref{fig:fig4}a, highly heated plasma flow
($\Theta \gtrsim 2$) is realized.

%---[ cut by MT ]
% In the previous sub-sections we explored that the shocked plasma
% flow can be strongly magnetized when $\tilde{L} \Omega_F$ is
% large.
%---

\begin{figure}[t]% ------------------------------------- Figure~7
    \centering
    \epsscale{0.5}
    \includegraphics[angle=0, width=5.1in]{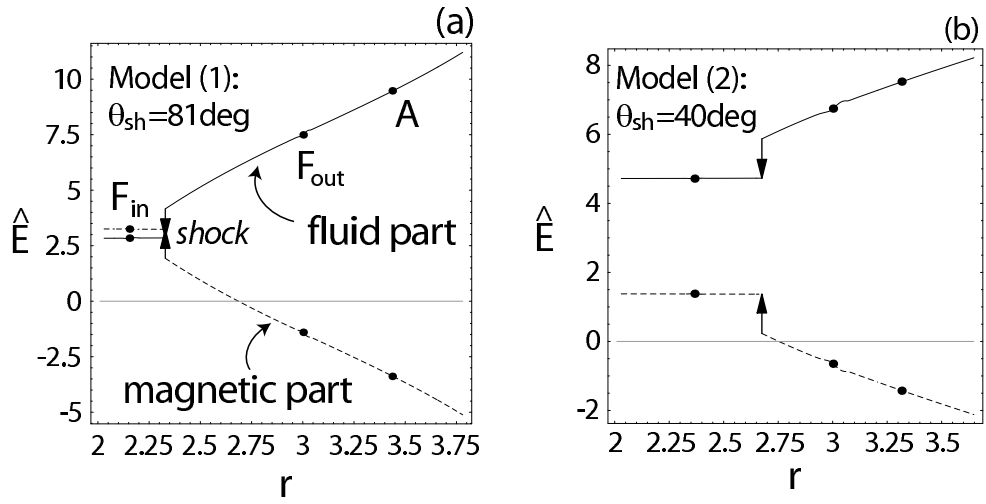}
    \caption{Radial profile of the energy distribution of $\hat{E}$ for
    models (1) and (2) in Figure~\ref{fig:fig2}b. (a) $\tilde{L} \Omega_F = 0.55$ for the model (1)
    and (b) $0.23$ for the model (2).
    See Table~\ref{tab:tbl-1} for the model parameters in detail. }
\label{fig:fig7}
\end{figure}

Next, we will examine the energy distribution of the plasma to
further understand the nature of the shocked plasma flows. In
Figure~\ref{fig:fig7} we show the components of energy of the
plasma, namely fluid $E_{\rm fluid}$ [first term of the
right-hand-side in equation~(\ref{eq:E})] and magnetic $E_{\rm
magnetic}$ [second term of the right-hand-side in
equation~(\ref{eq:E})] components, for models (1) and (2) in
Figure~\ref{fig:fig2}b. In the course of accretion the energy
transport between the fluid ($E_{\rm fluid}$) and the magnetic field
($E_{\rm magnetic}$) occurs. \citet{Hirotani92} discussed the energy
transport (or energy redistribution of the plasma) between the two
components in shock-free accreting MHD plasmas. In a slowly-rotating
black hole case ($0 \le \omega_H < \Omega_F$), the dominant
component in $E$ is fluid energy (i.e., $E_{\rm fluid} > E_{\rm
magnetic}$), but the magnetic component can become comparable to or
exceed the fluid component near the event horizon for lager
$\tilde{L} \Omega_F$-value. [see the model (1)]. Note that after the
shock each component almost remains unchanged. In other words, the
shock acts in such a way that this energy transport from the fluid
to the magnetic field is even enhanced at the shock front.

\section{Discussion}
In order to explore a general trend for MHD shock formation for the
vast amount of parameter sets, we sliced the entire parameter space
which consists of all of the parameters ($E,L,\Omega_F,\eta,r_{\rm
sh},\theta_{\rm sh};a,\delta$), by finding the allowed {\it cubic}
parameter space consisting of three parameters ($E,L,\Omega_F$), for
given black hole magnetosphere models with the other two parameters
$a$ and $\delta$ fixed. Since it is very hard to explore all the
parameter space spanned by all of the parameters together, our
current approach greatly simplifies and yet help us better
understand a comprehensive picture of the resulting shock solutions.
Through our search for possible non-equatorial MHD shocks we find
that the allowed shock region is constrained by various physical
factors - especially the regularity conditions at the magnetosonic
points and the shock conditions. Because the mathematical
expressions for the regularity conditions for the existence of the
\Alfven~ and fast magnetosonic points are not at all simple
\citep[e.g.,][]{Takahashi02}, the topological appearance of the
obtained shock regions (see, e.g., Figure~\ref{fig:fig3}) is
complicated, but nevertheless we find it very useful.

We find in general that stronger MHD shocks can form when the plasma
energy is larger. With magnetosphere rotation the shock becomes
stronger. The shocked plasma temperature has a clear
anti-correlation with the shocked plasma magnetization. The more
strongly magnetized plasma is formed for larger $\tilde{L}
\Omega_F$. We also find that the energy transport between the fluid
and the magnetic field can operate even more effectively across the
shock front. This transition is internally caused by the transport
between these components. Near the equatorial regions the shock is
generally stronger. It is thus possible for the astrophysical
accreting plasma that very powerful MHD shock formation takes place
near the the equatorial event horizon. In the present calculations
non-equatorial MHD shocks are possible up to high latitudes with
$\theta_{\rm sh}$ as small as $\sim 25\degr$. We found in separate
calculations that $\theta_{\rm sh}$ can be as small as $\sim 5\degr$
when $\Omega_F$ is sufficiently large. If the poloidal magnetic
field or resulting shock is dynamically unstable and switched on and
off as the rotating plasma continues to fall onto the event horizon,
some sort of quasi-periodic phenomena, perhaps with the dynamical
frequency of $\Omega(r_{\rm sh})$ may be expected (where $\Omega
\equiv u^\phi / u^t$ is the angular frequency of the shocked fluid).
On the other hand, particles (primarily thermal electrons) would be
accelerated at the shock front through first-order Fermi mechanism
\citep[]{Fermi49,Baring97,Gieseler00} in the presence of the
randomly-distributed, turbulent magnetic field. The shock front
could then be partly responsible for generating the base of the
relativistic winds/jets \citep[e.g.,][for the stellar wind
case]{Quataert05}.

When considering ingoing flows, we adopt mostly a split-monopole
poloidal field geometry near the horizon \citep[e.g.,][]{BZ77}, to
mimic the radial magnetic field lines very close to the horizon
(see Figure 1). Based on the ideal MHD conditions, the ingoing
plasma is assumed to be frozen-in to the magnetic field lines. The
ingoing plasma we considered in this work needs to be injected
from some plasma sources. Although we do not consider the exact
origin of such sources, there are some possible sites for
generating the ingoing plasma; e.g., the surface of the accretion
disk and/or its corona. In this case, the magnetic field lines
connecting to the black hole may come from the outer plasma
sources at large radii. Although we consider the conical field
lines near the horizon, the poloidal field line can also be bent
toward the equator and be connected to the equatorial accretion
disk at regions further away from the black hole. We expect that
such regions are at least outside the potential well of the
particle's effective potential. Note that in the description of
the steady-state black hole magnetosphere, the global poloidal
field coupling the black hole event horizon to the surface of a
geometrically-thin disk has already been considered (e.g., see
Nitta, Takahashi, \& Tomimatsu 1991; Tomimatsu \& Takahashi 1991
for the analytic solutions, and Komissarov 2005; Uzdensky 2005 for
the numerical results). The poloidal field configuration adopted
in our model is based on these studies.

In this paper, a typical value for the angular velocity of the field
line $\Omega_F$ which allows MHD shock formation is about twice that
of the Keplerian disk $\Omega_{\rm Kep}(r_{\rm ms})$ (where $r_{\rm
ms}$ is the the innermost radius). That is, if the black hole-disk
connecting field lines are considered under the conventional
Keplerian motion, adiabatic standing MHD shock formation is less
likely to occur. Our results suggest that the shock may develop if
the rotation of the magnetic field is super-Keplerian [i.e.,
$\Omega_F
> \Omega_{\rm Kep}(r_{\rm ms})$] due to some processes.
Although we may speculate on what this additional process may be,
that is beyond the scope of our current paper. \citet{McKinney04}
compared near-equatorial stationary MHD inflows with their
time-dependent numerical results between the innermost stable
circular orbit and the event horizon. In their inflow solution,
there is no MHD shock when the time-averaged numerical value of
$\Omega_F$ is about $10\%$ of the $\Omega_{\rm Kep}(r_{\rm ms})$.
Their result is consistent with our condition for the MHD shock
formation; that is, no shock solution exists for $\Omega_F \lesssim
\Omega_{\rm Kep}(r_{\rm ms})$.

We now show that the values of various primary parameters we chose
are not arbitrary, but they are chosen with possible applications to
the central engines of AGNs in mind. For instance, for a typical
magnetic field strength of AGNs, \citet[][]{Krolik99} estimates the
magnetic field strength as $B \sim 4 \times 10^3 \times T^2_5$ G
under the equipartition assumption where $T_5$ is the effective
temperature of the disk in units of $10^5$ K. For radio-loud
narrow-line quasars, \citet{Wang01} suggests that the field strength
of at least $B \sim 10^4$ G is required for magnetic-heated corona
to operate. Following their implications, we took a likely value of
$B_p \sim 10^4$ G for a typical accretion-powered Seyfert nuclei
($M_{BH} \sim 10^7 M_\odot$) in our estimate. Some of the primary
field-aligned parameters are $\tilde{L}$ and $\hat{\eta}$. The
physical value for the specific angular momentum $\tilde{L}$ must
lie in a certain range in such a way that the \Alfven~points do
exist (see Paper I for details). Since we consider the
trans-magnetosonic flows, our choice of $\tilde{L}$ should be
reasonable. $\eta$ tells us the particle flux per magnetic flux.
Using a commonly used value for the mass-accretion rate for Seyfert
nuclei (from observational estimates), $\dot{M}_{acc} \sim 10^{-2}
M_\odot$/year $\sim 10^{24}$ gram/sec, we can estimate
$\hat{\eta}$-value as $\hat{\eta} \sim \dot{M} r c^2/C^2$ (in
physical units) at the horizon. Note that we only take 1\% of
$10^{24}$ gram/sec for "our" non-equatorial MHD plasma flows. That
is, $\dot{M}_{plasma} \sim 0.01 \dot{M}_{acc} \sim 10^{22}$
gram/sec. This is because the plasma density of the non-equatorial
flows should be considerably lower than that of the equatorial
flows. Also, we have $B_p \sim C/r^2$, $\eta \equiv n r u^r c/B_p$,
and $\dot{M}_{plasma} \sim r^2 u^r n m_p$. By eliminating $u^r, m_p,
\eta$ from these equations, we get $\hat{\eta} \sim r
\dot{M}_{plasma} c^2/C^2 \sim 0.01$, which is in the range of values
adopted in our calculations. Note also that the estimated value of
$\hat{\eta}$ depends on the values of the black hole mass and the
magnetic field strength. Depending on these values,
$\hat{\eta}$-value can be smaller or larger than $0.01$ by orders of
magnitude.
%For particle number flux per magnetic flux $\hat{\eta}$-value, we
%estimate $\hat{\eta} \sim \dot{M} r_H / C^2 \sim 0.01$, where we
%have assumed the mass-accretion rate for our non-equatorial flow to
%be $\dot{M} \sim 10^{-4}$ M$_\odot$/year $\sim 10^{22}$ grams/sec
%(\textbf{based on Seyfert observations}). \textbf{Note that in this
%application a small fraction (about $1\%$) of the total accretion
%rate for typical Seyferts applies to the non-equatorial MHD flow
%which we consider in this paper. This is because the plasma density
%of the non-equatorial MHD flows should be considerably lower than
%that of the equatorial flows.} Thus, the $\hat{\eta}$-values we
%adopted in our calculations are reasonable. \textbf{Note also that
%the estimated value of $\hat{\eta}$ depends on the values of the
%black hole mass and the magnetic field strength. Depending on these
%values, $\hat{\eta}$-value can be smaller or larger than 0.01 by
%orders of magnitude.}

Also, in an attractive model of accretion-powered AGNs, the critical
outer region of the disk where the inflowing plasma originates is
also where the outflowing plasma is injected. For the outgoing
flows, the Lorentz factor $\gamma$ of the observed relativistic
jets/winds is generally $\hat{E}=\hat{E}_\infty \sim \gamma \sim 10$
(for microquasar black hole systems) to $\gamma \gtrsim 100$ (AGNs
and Gamma Ray Bursts (GRBs)) \citep[e.g.,][]{Meier03}. That implies
the presence of highly energetic plasmas being accelerated within
the black hole magnetosphere. Because they are injected to form the
jets/winds with such high observed energy, $\hat{E}_\infty$, the
ingoing plasmas also should initially possess the same order of
magnitude of energy $\hat{E} \sim \hat{E}_\infty$ at the time of
their launch from the foot points on the accretion disk. Therefore,
the original energy of the MHD flow can also be $\hat{E} \sim
10-100$\footnote[1]{For a given $\tilde{L} \Omega_F$-value, the
$\hat{\eta} \hat{E}$-value is nearly constant (see Paper I). In the
present work we do find MHD shocks for $\hat{E} \sim 10$ with
$\hat{\eta} \sim 10^{-3}$ and $\hat{E} \sim 100$ with $\hat{\eta}
\sim 10^{-4}$.}. A major part of the energy can be magnetic in this
case near the plasma source on the disk. Our selected value for
$\hat{E}$ ($\hat{E}>1$) in our calculations is then justified. For
completeness, we would like to mention that no plasma energetically
bound to the black hole (i.e., $\hat{E}\lesssim 1$) appears to get
shocked in our case.

%\textbf{Adiabatic jump conditions are valid when the cooling process
%at the shock front is very inefficient. Otherwise, the heat
%generated at the shock front would be quickly dissipated from the
%plasma flow (radiatively and/or kinematically), causing the
%downstream temperature to be continuous across the shock front.
%Another possibility under high conductive processes would be the
%formation of isothermal shocks in accretion and outflows
%\citep[e.g.,][for observational evidences of these types of
%shocks]{Forman05,Fabian06}.}

We are aware of the degeneracy of our solutions, i.e., the
coexistence of shock-free and shock-included solutions for the same
set of conserved quantities. To further investigate which
steady-state solution between the two is physically accessible in
nature, we will need to perform a dynamical analysis by considering
a time-dependent perturbation in the flow. Such an analysis is
beyond the scope of this paper. However, we may note that some
authors \citep[e.g.,][]{Ferrari84,Trussoni88,Nobuta94} have already
carried out such an analysis (no magnetic field in a flat space) and
concluded that {\it some shock-included solutions are indeed
preferred} (physically accessible by nature) to the corresponding
shock-free solution. That finding probably justifies the motivations
of our current work.

%%In this sense their work partially assures the credibility
%o%f our shock solutions - too strong!!!

Our current investigations carried out through the analytical
methods can help gain deeper insight when compared with the
results obtained by long dynamical MHD simulations. For instance,
when the numerical experiments for black hole magnetospheres are
studied under various initial conditions, the MHD shock front may
be generated in some computational domain after a long calculation
time to a quasi-steady state. Then, we can compare the
field-aligned flow parameters for MHD shock formation in our
analytic studies with the final values of the corresponding
physical quantities obtained by the numerical simulations.

Recently, long-time dynamical evolution of magnetized plasma
accretion onto a black hole has been extensively studied, and the
global poloidal magnetic fields are obtained by some authors
\citep[e.g.,][]{Hirose04,McKinney04,DeVilliers05,McKinney06}. The
magnetic field confined in the funnel region is evolved to
quasi-steady states at later times, while the magnetic field in the
corona and disk fluctuates in magnitude and its direction. Our
results may apply to some regions during such a quasi-steady-state
in late simulation times. However, the accreting flows (whether
equatorial or non-equatorial) are expected to be turbulent, and
hence our analysis here, based on axisymmetry and stationary
assumptions, is not exactly applicable for explaining short
timescale local turbulence in these regions. With this in mind, it
is still important to compare our results with the future
large-scale MHD simulations in order to better understand the
physics in these complex regions.

\section{Summary and Concluding Remarks}
In Paper I we showed that the MHD shock formation can occur in the
equator. In the present work we extended that work to the
non-equatorial two-dimensional geometry. We systematically
explored, for the first time, general relativistic MHD shock
formation in such a geometry by employing the conserved
quantities, and found the allowed shock regions. In summary we
find that non-equatorial MHD shocks can form and could be a
plausible candidate for generating a hot or strongly magnetized
region over various latitudes in the non-equatorial plane.

In various astrophysical objects, high energy activities are often
associated with the magnetic fields along which the accreting
plasma flows. Our investigation of MHD standing shock formation
will be useful for our better understanding of complicated
interactions between the plasma and the magnetic field in these
situations. In the context of a strong radiation source required
for the accretion-powered central engines of objects such as AGNs
and GBHCs, the presence of non-equatorial shocks therefore can be
attractive as a candidate for such a source, and therefore our
current work may turn out to be very interesting for future
observations of these objects also.

Before closing, it may be emphasized that our search for the
parameter space which allows MHD shock formation is carried out in
a systematic manner in such a way that {\it our choice of the
parameter sets is not arbitrary}. It is consistent, for instance,
with the numbers relevant in application to, e.g., the environment
of black hole magnetospheres around supermassive black holes in
AGNs. Within this context, our results are quite general. However,
it may be noted that in reality the MHD shock location in the
actual astrophysical situation would probably trace a certain
trajectory, rather than occupying the whole allowed shock region
in the constrained parameter space, because the parameters
allowing the shock formation may be unique from case to case. Our
hope is that our current studies, through analytic and
steady-state analysis, are very useful, because they can be
complementary to those with time-dependent numerical simulations,
in the sense that the latter can take advantage of, for instance,
the constrained parameter space found with our current studies.

\acknowledgments

The authors thank the anonymous referee for constructive criticism
to improve the manuscript. KF is grateful to A. Liebmann and R.
Takahashi for their fruitful comments, and K. Ohsuga for his
stimulating discussion. This work was supported in part by the
Grants-in-Aid of the Ministry of Education, Culture, Sports, Science
and Technology of Japan (17030006,M.T.).

\clearpage

\begin{deluxetable}{cccccccccc}%[h]% ------------------------------- Table~1
\tabletypesize{\scriptsize} \tablecaption{Characteristics of models
(1), (2) and (3) with different $(\hat{E},\tilde{L})$ in
Figure~\ref{fig:fig2}. $a=0, \delta=0, \Omega_F=0.14$ and
$\hat{\eta}=0.006$ for all cases. \label{tab:tbl-1}}
\tablewidth{0pt} \tablehead{ Model & Symbol & $\hat{E}$ &
$\tilde{L}$ & $\theta_{\rm sh}$ & $\lambda$ & $r_{\rm sh}$ &
$\Theta$ & $\sigma_2$ & $\dot{\cal{M}}_2$} \startdata
          (1) & $\bigcirc$ & 6.1 & 3.9  & $81\degr$ & 1.41 & 2.3 &  0.46 & 1.13 & 1.31  \\
          (2) & $\square$ & 6.1 & 1.6  & $40\degr$ & 1.15 & 2.7 & 0.29 & 0.29 & 1.51 \\
          (3) & $\bigtriangleup$ & 7.2 & 3.9  & $77\degr$ & 1.91 & 2.6 & 2.12 & 0.45 & 10.2 \\
\enddata
\end{deluxetable}

 %-----------------------------------------------

 %-----------------------------------------------

 %-----------------------------------------------

 %-----------------------------------------------

 %-----------------------------------------------

 %-----------------------------------------------

 %-----------------------------------------------


\begin{thebibliography}{99}


\bibitem[Appl \& Camenzind(1988)]{AC88} Appl, S., \& Camenzind, M., 1988, \aap, 206, 258 (AC88)
%\bibitem[Ballantyne, Turner, \& Young(2005)]{Ballantyne05} Ballantyne, D. R., Turner, N. J., \& Young, A. J.
%2005, \apj, 619, 1028
\bibitem[Baring(1997)]{Baring97} Baring, M. G. 1997, astro-ph/9711177
\bibitem[Bekenstein \& Oron(1978)]{Bekenstein78} Bekenstein, J. D., \& Oron, E. 1978, Phys. Rev. D, 18, 1809
%\bibitem[Beskin(1997)]{Beskin97} Beskin, V. S. 1997, Phys. Uspekhi 40, 659
\bibitem[Blandford \& Znajek(1977)]{BZ77} Blandford, R. D. \&
Znajek, R. L. 1977, \mnras, 179, 433
%\bibitem[Breitmoser \& Camenzind(2000)]{Camenzind00} Breitmoser, E., \& Camenzind, M. 2000, \aap, 361, 207
\bibitem[Camenzind(1986)]{Camenzind86} Camenzind, M. 1986, \aap, 162, 32

\bibitem[Camenzind(1987)]{Camenzind87} Camenzind, M. 1987, \aap, 184, 341
\bibitem[Chakrabarti(1990)]{Cha90} Chakrabarti, S. K. 1990, Theory of Transonic Astrophysical Flows (World Scientific,
Singapore)
%\bibitem[Das(2000)]{Das00} Das, T. K. 2000, MNRAS, 318, 294
%\bibitem[Das, Rao, \& Vadawale(2003)]{Das03} Das, T. K., Rao, A. R., \& Vadawale, S. V. 2003, \mnras, 343, 443
\bibitem[De Sterck \& Poedts(2000)]{DeSterck00} De Sterck, H.,
Poedts, S. 2000, PhRvL, 84, 5524
\bibitem[De Sterck \& Poedts(2001)]{DeSterck01} De Sterck, H.,
Poedts, S. 2001, Journal of Geophysical Research, 106, 12, 5524

\bibitem[De Villiers et al.(2005)]{DeVilliers05} De Villiers, J. P., Hawley, J. F., Krolik, J. H., \&
Hirose, S. 2005, \apj, 620, 878
%\bibitem[Di Matteo, Esin, \& Fabian(1999)]{Matteo99} Di Matteo, T., Esin, A., Fabian, A. C., \& Narayan,
%R. 1999, \mnras, 305, L1
%\bibitem[Fabian et al.(2000)]{Fabian00} Fabian, A. C., Iwasawa, K., Reynolds, C. S., \& Young, A. J. 2000, \pasp, 112, 1145
\bibitem[Fabian et al.(2002)]{Fab02} Fabian, A. C., Vaughan, S., Nandra, K., Iwasawa, K., Ballantyne, D. R., Lee, J. C.,
De Rosa, A., Turner, A., \& Young, A. 2002, \mnras, 335, L1

%\bibitem[Fabian et al.(2003)]{Fabian03} Fabian, A. C., Sanders, J. S., Allen, S. W., Crawford, C.
%S., Iwasawa, K., Johnstone, R. M., Schmidt, R. W., \& Taylor, G.
%B. 2003, \mnras, 344L, 43F
%\bibitem[Fabian et al.(2006)]{Fabian06} Fabian, A. C., Sanders, J. S., Taylor, G. B., Allen, S.
%W., Crawford, C. S., Johnstone, R. M., \& Iwasawa, K. 2006,
%\mnras, 366, 417F
%\bibitem[Fabian et al.(2005)]{Fabian05} Fabian, A. C., Miniutti, G., Iwasawa, K., \&
%Ross, R. R. 2005, \mnras, 361, 795

\bibitem[Fermi(1949)]{Fermi49} Fermi, E. 1949, Phys. Rev. Lett. 75, 1169
\bibitem[Ferrari et al.(1984)]{Ferrari84} Ferrari, A., Habbal, S. R., Rosner, R., \& Tsinganos, K. 1984, \apj, 277, L35

%\bibitem[Forman et al.(2005)]{Forman05} Forman, W., Nulsen, P., Heinz, S., Owen, F., Eilek, J., Vikhlinin,
%A., Markevitch, M., Kraft, R., Churazov, E., \& Jones, C. 2005,
%\apj, 635, 894

\bibitem[Fukumura \& Tsuruta(2004)]{FT04} Fukumura, K., \&
Tsuruta, S. 2004, \apj, 611, 964 (FT04)
\bibitem[Fukumura(2005)]{Fukumura05} Fukumura, K. 2005, Ph.D. Thesis, Montana State
University

%\bibitem[Goto(2003)]{Goto03} Goto, J. 2003, Master Thesis, Aichi University
%of Education

\bibitem[Gammie, Shapiro, \& McKinney(2004)]{Gammie04} Gammie, C.
F., Shapiro, S. L., \& McKinney, J. C. 2004, \apj, 602, 312
\bibitem[Gieseler \& Jones(2000)]{Gieseler00} Gieseler, U. D. J., \& Jones, T. W. 2000, \aap, 357,
1133

%\bibitem[Grad \& Rubin(1958)]{Grad58} Grad, H., \& Rubin, H. 1958, International Atomic Energy Agency
%Conf. Proc. 31 (Geneva), p.190

\bibitem[Haardt \& Maraschi(1991)]{Haardt91} Haardt, F. \& Maraschi, L. 1991, \apj, 380, L51
\bibitem[Hada(1994)]{Hada94} Hada, T. 1994, Geophys. Res. Lett., 21, 2275
\bibitem[Hirose et al.(2004)]{Hirose04} Hirose, S., Krolik, J. H., De Villiers, J.-P., \&
Hawley, J. F. 2004, \apj, 606, 1083

\bibitem[Hirotani et al.(1992)]{Hirotani92} Hirotani, K., Takahashi, M., Nitta, S., \& Tomimatsu, A., 1992, \apj, 386, 455

\bibitem[Iwasawa et al.(1996a)]{Iwasawa96a} Iwasawa, K., Fabian, A. C., Mushotzky, R. F., Brandt, W. N., Awaki, H.,
\& Kunieda, H. 1996a, \mnras, 279, 837
\bibitem[Iwasawa et al.(1996b)]{Iwasawa96b} Iwasawa, K., Fabian, A. C., Reynolds, C. S., Nandra, K., Otani, C., Inoue, H.,
Hayashida, K., Brandt, W. N., Dotani, T., Kunieda, H., Matsuoka,
M., \& Tanaka, Y. 1996b, \mnras, 282, 1038
\bibitem[Koide et al.(2000)]{Koide00} Koide, S., Meier, D. L., Shibata, K., \& Kudoh, T.
2000, \apj, 536, 668
\bibitem[Komissarov(2005)]{Komissarov05} Komissarov, S. S. 2005, \mnras, 359, 801
\bibitem[Krolik (1999)]{Krolik99} Krolik, J. H. 1999, Active Galactic Nuclei (Princeton
University Press, New Jersey)

%\bibitem[(Kusunose \& Mineshige(1995))]{Kusunose95} abc...
%\bibitem[Le \& Becker(2004)]{Le04} Le, T. \& Becker, P. A. 2004, \apj, 617, L25
%\bibitem[Li(2002a)]{Li02a} Li, L.-X. 2002a, \apj, 567, 463

\bibitem[Li(2002)]{Li02} Li, L.-X. 2002, Phys. Rev. D. 65, 084047

%\bibitem[Liu, Mineshige, \& Shibata(2002)]{Liu02} Liu, B. F., Mineshige, S., \& Shibata K. 2002,
%\apj, 572, L173
%\bibitem[Liu, Mineshige, \& Ohsuga(2003)]{Liu03} Liu, B. F., Mineshige, S., \& Ohsuga, K. 2003, \apj,
%587, 571

\bibitem[Lu et al.(1997)]{Lu97} Lu, J.-F., Yu, K. N., Yuan, F., \& Young, E. C. M. 1997, \aap, 321, 665
\bibitem[Lu \& Yuan(1998)]{Lu98} Lu, J.-F., \& Yuan, F. 1998, \mnras, 295, 66

%\bibitem[Machida, Hayashi, \& Matsumoto(2000)]{Machida00} Machida, M., Hayashi, Mitsuru R., Matsumoto, R.
%2000, \apj, 532, L67

\bibitem[McKinney \& Gammie(2004)]{McKinney04} McKinney, J. C. \& Gammie, C. F. 2004, \apj, 611, 977

\bibitem[McKinney(2006)]{McKinney06} McKinney, J. C. 2006,
\mnras, 368, 1561

\bibitem[Meier(2003)]{Meier03} Meier, D. L. 2003, New Astronomy
Reviews, 47, 667
\bibitem[Meier(2004)]{Meier04} Meier, D. L. 2004, preprint (astro-ph/0504511)

%\bibitem[Michel(1973a)]{Michel73a} Michel, F. C. 1973a, \apj, 180, 207
%\bibitem[Michel(1973b)]{Michel73b} Michel, F. C. 1973b, \apj, 180, L133
%\bibitem[Miniutti \& Fabian(2004)]{Fabian04} Miniutti, G., \& Fabian, A. C. 2004,
%\mnras, 349, 1435

\bibitem[Mobarry \& Lovelace(1986)]{Mobarry86} Mobarry, C. M., \& Lovelace, R. V. E. 1986, \apj,
309, 455
\bibitem[Nandra \& Pounds(1994)]{Nandra94} Nandra, K., \& Pounds, K. A. 1994, \mnras,
268, 405
\bibitem[Nitta, Takahashi, \& Tomimatsu(1991)]{NTT91} Nitta, S., Takahashi, M., \& Tomimatsu, A. 1991, Phys.Rev.D 44,
2295
\bibitem[Nobuta \& Hanawa(1994)]{Nobuta94} Nobuta, K., \& Hanawa, T. 1994, PASJ, 46, 257
\bibitem[Pounds et al.(1994)]{Pounds94} Pounds, K., Nandra, K., Fink, H. H., \&
Makino, F. 1994, \mnras, 267, 193
\bibitem[Quataert \& Loeb(2005)]{Quataert05} Quataert, E., \& Loeb, A. 2005, \apj,
635, L45
%\bibitem[Punsly(2001)]{Punsly01} Punsly, B. 2001, Black Hole Gravitohydromagnetics (Springer)
%\bibitem[Reynolds \& Nowak(2003)]{Reynolds03} Reynolds, C. S., \& Nowak, M. A. 2003, Physics Report, 377,389
\bibitem[Rilett(2003)]{Rilett03} Rilett, J. D. 2003, Ph.D. Thesis, Montana State
University (R03)
%\bibitem[Shafranov(1966)]{Shafranov66} Shafranov, V. D. 1966, in Reviews of Plasma Physics, ed. M. A.
%Leontovich (New York: Consultants Burear), 2, 103

%\bibitem[Shibata, Tajima, \& Matsumoto(1990)]{Shibata90} Shibata, K., Tajima, T., \& Matsumoto, R. 1990, \apj,
%350, 295

%\bibitem[Sterling \& Hudson(1997)]{Sterling97} Sterling, A. C., \& Hudson, H. S. 1997, \apj, 491, L55

\bibitem[Takahashi et al.(1990)]{TNTT90} Takahashi, M., Nitta, S., Tatematsu, Y., \& Tomimatsu, A. 1990, \apj, 363, 206
\bibitem[Takahashi et al.(2002)]{TRFT02} Takahashi, M., Rilett, D., Fukumura, K., \& Tsuruta, S. 2002, \apj, 572,
950 (TRFT02)
\bibitem[Takahashi(2002)]{Takahashi02} Takahashi, M. 2002, \apj, 570, 264

\bibitem[Takahashi et al.(2006)]{Takahashi06} Takahashi, M., Goto,
J., Fukumura, K., Rillet, D., \& Tsuruta, S. 2006, \apj, 645, 1408
(Paper I)

\bibitem[Tomimatsu \& Takahashi(2001)]{TT01} Tomimatsu, A., \& Takahashi, M. 2001, \apj, 552, 710
\bibitem[Trussoni et al.(1988)]{Trussoni88} Trussoni, E., Ferrari, A., Rosner, R., \& Tsinganos,
K. 1988, \apj, 325, 417
\bibitem[Uzdensky(2005)]{Uzdensky05} Uzdensky, D. A. 2005, \apj, 620, 889
\bibitem[Wang et al.(2001)]{Wang01} Wang, T. G., Matsuoka, M., Kubo, H., Mihara, T., \&
Negoro, H. 2001, \apj, 554, 233
\bibitem[Wilms et al.(2001)]{Wilms01} Wilms, J., Reynolds, C. S., Begelman, M. C., Reeves, J., Molendi, S., Staubert, R.,
\& Kendziorra, E. 2001, \mnras, 328, L27


\end{thebibliography}
\end{document}